\documentclass[prb,nofootinbib,amsfonts,tightenlines,twocolumn]{revtex4-1}

\usepackage{graphicx}
\usepackage{dcolumn}
\usepackage{bm}
\usepackage{caption}
\usepackage{subcaption}
\usepackage{mathrsfs}
\usepackage{dsfont}
\usepackage{physics}
\usepackage{natbib}
\usepackage{xcolor}
\usepackage{wrapfig}\usepackage{upgreek}
\usepackage[utf8]{inputenc}
\usepackage{xpatch}

\usepackage{soul}

\appendix

\begin{document}

\title{Thermal signature of  Majorana fermions in Josephson junction
}

\author{Aabir Mukhopadhyay} 
\email{aabir.riku@gmail.com}
\author{Sourin Das }
\email{sourin@iiserkol.ac.in, \\ ORCID ID : 0000-0002-8511-5709}
\affiliation{Indian Institute of Science Education \& Research Kolkata,
Mohanpur, Nadia - 741 246, 
West Bengal, India}
\date{\today}

\begin{abstract}
Existence of distinctive correlation between the thermal conductance and the Josephson current across a 1-D topological Josephson junction (T-JJ) is established and an expression connecting the two is derived in the short junction limit. It is shown that the three terminal T-JJ provides us with an ideal set-up for exploiting the distinctive correlations which aid in identifying hallmarks of Majorana bound state localized at the junction. Our approach combines information form sub-gap Cooper pair transport and transport of above-gap thermally excited quasiparticle which is complementary to the existing studies.
\end{abstract}

\maketitle
%%%%%%%%%%%%%%%%%%%%%%%%%%%%%%%%%%%%%%%%%%%%%%%%%%%%%%%%%%%%%%%%%%%%%%%%%%%%%%%%%%%%%%%%%%%%%%%%%%%%%%%%%%%%%%%%%%%%%%%%%%
{\underline{\it{Introduction}}}:
Maki and Griffin\cite{maki_PRL_15_921, maki_PRL_16_258}  carried out theoretical study of heat transport across a Josephson junction (JJ) in 1965 and ever since there have been several studies pertaining to  thermal transport across JJs\cite{guttman_PRB_55_3849, zhao_PRL_91_077003, zhao_PRB_69_134503, fornieri_NN_12_425, luo_SR_9_2187} including some recent developments involving two and three dimensional topological insulators\cite{sothmann_PRB_94_081407, sothmann_NJP_19_023056}.  On the experimental side, the theoretical predictions\cite{maki_PRL_15_921, maki_PRL_16_258} were confirmed in a remarkable experiment by
 Giazotto and Martinez-Perez in 2012\cite{giazotto_Nat_492_401}
  %
%Owing to remarkable progress in experimental study of thermally biased JJ, the theoretical predictions of Maki and Griffin has been confirmed by Giazotto and Martinez-Perez\cite{giazotto_Nat_492_401} in 2012 and
%
which led to steady experimental progress in this field\cite{golubev_PRB_87_094522, martinez_APL_102_182602, martinez_JLTP_175_813, martinez_NC_5_3579, martinez_NN_10_303, fornieri_NN_11_258, kolenda_PRL_116_097001,fornieri_PRB_93_134508,  tan_NC_8_15189, fornieri_NN_12_944, paolucci_PRA_10_024003, karimi_PRA_10_054048}.\\
Proximity induced superconductivity in spin-orbit nano-wire and the helical edge state (HES) of quantum spin Hall (QSH) state allow for two distinct realizations 
of topological superconductivity and localized Majorana bound states (MBS)\cite{lutchyn_PRL_105_077001, oreg_PRL_105_177002, nadj_Nat_468_1084, liang_NL_12_3263, mourik_Sc_336_1003, anindya_NP_8_887, sestoft_PRM_2_044202, bommer_PRL_112_187702, konig_Sc_318_766, brune_NP_6_448, knez_PRL_107_136603, brune_NP_8_485, knez_PRL_109_186603, hart_NP_10_638}. In case of a nano-wire, these MBS are expected to appear at the two ends of the wire\cite{Kitaev_2001}  while for Helical edge state (HES) of QSE state, localized MBS forms at the junction of a proximity induced superconducting region and a region exposed to  Zeeman field\cite{fu_PRB_79_161408,PhysRevLett.101.120403}.  It is pertinent to note that experimental realization of T-JJs\footnote{Topological JJ refers to those JJ where the sub-gap bound state formed at the junction are primarily due to overlap of Majorana bound state appearing at the ends of the topological supercondcutors.} has been achieved both in the context of nano-wire set-up\cite{laroche_NC_10_245}  and HEH of QSE state set-up\cite{bocquillon_NN_12_137}. These experimental advances call for a renewed look at  how to exploit a combination of electrical and thermal means to detect and manipulate the MBS.

\begin{figure}
%\centering
\includegraphics[scale=0.6]{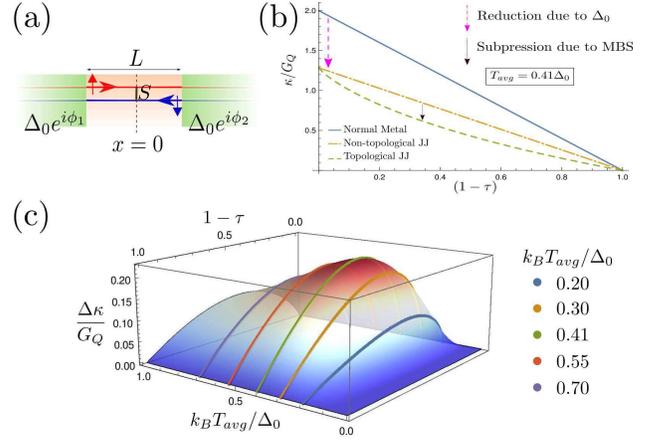}
\caption{(a) Pictorial representation of two-terminal JJ hosted on HES where a backscatterer $S$ is introduced at $x=0$. (b) Plot of thermal conductance $\kappa$ for $\phi=0$ in units of $G_Q$ is plotted as a function of $\tau$ for a normal junction (non-superconducting) and the corresponding non-topological and topological JJ at $\phi=0$. (c) Three dimensional surface plot of difference between topological and non-topological case  in the $(1$-$\tau)$-$ (k_B T_{\text{avg}}/\Delta_0)$ plane. 
%The coloured lines are contours of fixed $(k_B T_{avg}/\Delta_0)$ where $(k_B T_{avg}/\Delta_0)=.41$ contour passes through the maximum value of $\Delta\kappa/G_Q$. 
}
\label{conductance_two_terminal}
\end{figure}

As far as detection  of MBS via Josephson effect is concerned, observation of $4 \pi$ Josephson effect\cite{kwon_TEPJB_37_349, fu_PRB_79_161408} is considered as a hallmark of MBS which primarily relates to sub-gap physics. 
 The central focus of this article is to establish that the  above-gap transport of thermally excited bulk quasi-particle ($QP$) across a JJ also carry distinctive signatures of topological superconductivity and MBS. This can be understood as follows. \\ 
It is known that the thermal current across a JJ in the weak tunneling limit can be expressed as a sum of contribution due to $QP$s tunneling, Cooper pairs tunneling (which is usually negligible\cite{pershoguba_PRB_99_134514})  and the interference of the two\cite{maki_PRL_15_921, maki_PRL_16_258, guttman_PRB_57_2717}. For a JJ with arbitrary transparency, these interferences simply lead to  zero in denominator of transmission probability of $QPs$ at energies (denoted by $\omega_0$) at which Andreev bound state (ABS)\cite{sauls_TRSP_2018, beenakker_PRL_67_3836, PhysRevB.27.6739, PhysRev.187.556} are formed in the JJ. Remarkably, the phase bias ($\phi$) of 1-D T-JJ enters the $QP$ transmission probability solely via the dependence of sub-gap bound states energy($\omega_0 (\phi)$) on $\phi$. In contrast, the $\phi$ dependence of a 1-D non-topological JJ appears also in the numerator of the $QP$ transmission probability resulting in a complicated function of $\phi$ (see Appendix A).  Recall that the Josephson current  is dominated by contributions from $\phi$ derivative of the sub-gap bound state energy, $\omega_0 (\phi)$ in the short junction limit\cite{beenakker_PRL_67_3836}. Hence, the $\phi$ dependence of both the heat current carried by the thermally excited $QP$ and Josephson current carried by Cooper pairs acquire their $\phi$ dependence primarily from their dependence on $\omega_0 (\phi)$, leading to a distinctive correlation between the two independently measurable quantities. An additional correlation which exists on a similar footing as $\phi$ arises via the dependence of sub-gap bound state energy ($\omega_0$) on the normal state transmission probability ($\tau$). In what follows, we will focus on the short JJ limit unless stated otherwise.  \\
We study correlation between Josephson current, thermal conductance and the corresponding normal state electrical conductance of a three terminal T-JJ as a function of ($\phi$,$\tau$). We derive a set of relations (see Eq.[\ref{K_with_s}] and  Eq.[\ref{D_K_with_t_and_Ij}]) between these quantities. The validity of these relations owes its existence to the presence of MBS at the junction and therefore is a smoking gun signature of MBS. We stress that the three terminal T-JJ set-up is expected to host a single MBS pinned at zero energy\footnote{This is so because each topological superconductor will contribute a single Majorana to the Josephson tri-junction but two of them will hybridize and gap out while the third MBS will stay topologically protected.} and therefore is of particular importance as a testing ground for MBS.  Note that a study involving thermal conductance of topological ABS was reported in Ref.[\onlinecite{sothmann_PRB_94_081407}] though it did not touch upon the ideas presented here, i.e., the detection of MBS using a multi-terminal set-up or correlation between electrical and thermal response of T-JJ. \\
\underline{\textit{Multi-terminal thermal conductance}}: For a multi-terminal junction, heat currents driven by temperature bias within linear response theory can be expressed as
\begin{equation}
I^h_n= \sum_{m\neq n} \kappa_{n,m} (T_n-T_m)	
\label{heat_current}
\end{equation}
where $\kappa_{n,m}$ is the thermal conductance between terminal $m$, $n$ and $T_n$, $T_m$ are the temperatures of terminal $n$ and $m$ respectively. We consider a general multi-terminal junction of superconducting leads (described by BdG Hamiltonian\cite{de_gennes_book}) connected via a common normal region. An incident electron-like quasiparticle in $m$th lead (with energy $\omega$) results in a  reflected electron-like and hole-like quasiparticle within the same lead with probabilities, say $\mathcal{R}^{m,m}_{e,e}$ and $\mathcal{R}^{m,m}_{h,e}$ respectively along with transmitted electron-like and hole-like quasiparticle in the $n$th lead $(n \neq m)$ with probabilities $\mathcal{T}^{n,m}_{e,e}$ and $\mathcal{T}^{n,m}_{h,e}$ respectively. Probability conservation ensures that  $(\mathcal{R}^{m,m}_{e,e}+\mathcal{R}^{m,m}_{h,e})+ \sum_{n} (\mathcal{T}^{n,m}_{e,e}+\mathcal{T}^{n,m}_{h,e})=1$. We denote $\mathcal{T}^{n,m}=\mathcal{T}^{n,m}_e+\mathcal{T}^{n,m}_h=(\mathcal{T}^{n,m}_{e,e}+\mathcal{T}^{n,m}_{h,e})+(\mathcal{T}^{n,m}_{h,h}+\mathcal{T}^{n,m}_{e,h})$. The multi-terminal linear response thermal conductance between leads $m$ and $n$ is then given by \cite{sothmann_PRB_94_081407}
\begin{equation}
\kappa_{n,m}(\phi_j)=\left[\, \dfrac{1}{h}\int_{\Delta_0}^{\infty} d\omega\, \omega \, \{ \mathcal{T}^{n,m} \} \dfrac{df(\omega,T)}{dT}\right]_{T=T_{\text{avg}}},\label{def_of_conductance}
\end{equation}
where  $f(\omega,T)$ is the Fermi distribution function at temperature $T$, $\Delta_0$ is the superconducting gap (taken to be same of all leads) and  $T_{\text{avg}}$ is the average junction temperature.\\
{\underline{\textit{Two Terminal T-JJ in HES: }}}
Though our findings are valid for 1-D T-JJ based on either  {\it(a)} 1-D electrons with quadratic dispersion and proximitized p-wave superconductivity\cite{oindrila_PRB_97_174518} or  {\it(b)} 1-D Dirac fermions with proximitized s-wave superconductivity,  we will primarily discuss the case of 1-D Dirac fermions realized in HES of QSH state. We consider a JJ in HES where the junction is defined by $|x|<L/2$ region and we have a finite pair potential $\Delta_0 e^{i \phi_r}$ ($[r \in \{ 1,2 \}]$) for $\infty>|x|>L/2$  (see Fig[\ref{conductance_two_terminal}] (a)).
Our set-up is described by Bogoliubov-de Gennes Hamiltonian\cite{fu_PRB_79_161408, dolcini_PRB_92_035428} in the Nambu basis ($\Psi=[(\psi_{\uparrow},\psi_{\downarrow}),(\psi_{\downarrow}^{\dagger},-\psi_{\uparrow}^{\dagger})]^T$\citep{fu_PRL_100_096407}) given by 
\begin{align}
\mathcal{H} = ( v_F \hat{p}_x \sigma_z-\mu)\uptau_z 
+ \Delta(x) (\cos \phi_r \uptau_x -\sin \phi_r \uptau_y),
\label{general_Hamiltonian}
\end{align}
where $\Delta(x)=\Delta_0 [\Theta(-x-L/2)+\Theta(x-L/2)]$,  $ \sigma$ and $\uptau$ represents spin and particle-hole degrees of freedom respectively.  We focus on the highly doped regime given by  the chemical potential $\mu>>\Delta_0$ and at the short junction limit given by $L<<\xi$, where $\xi=\hbar v_F/\Delta_0$ is the superconducting coherence length. For $\omega<\Delta_0$ (measured with respect to $\mu$) Andreev bound states (ABS) are formed at energy $\omega_0^{\pm}=\pm \Delta_0 \sqrt{\tau} \cos \phi/2$ $(\phi=\phi_2-\phi_1)$ (see Supplemental Material: Section A) while for $\omega>\Delta_0$ we have propagating quasiparticle solutions. Here $\tau$ represents transmission probability scattering matrix $S$ introduced at  $x=0$ and is assumed to be energy independent. The quasiparticle transmission probability across the JJ described by Eq.[\ref{general_Hamiltonian}],  
$\mathcal{T}^{1,2}$, can be expressed as (see Supplemental Material: Section A)
\begin{equation}
\mathcal{T}^{1,2} (\omega,\tau,\phi) = \tau ~ 
\left[\dfrac{ 2\, (\omega^2-\Delta_0^2) \, \mathfrak{f}(\omega,\tau,\phi)}{(\omega^2 -\vert \omega_0^{\pm} (\tau,\phi) \vert ^2
)^2}\right]
\,
\Theta(\vert \omega \vert^2-\Delta_0^2) \,,
\label{4}
\end{equation}
\noindent which is general form for $\mathcal{T}$ in case of 1-D JJ in the short junction limit. For the T-JJ, $\mathfrak{f}(\omega,\tau,\phi)= (\omega^2-\vert \omega_0^{\pm}(\tau,\phi) \vert^2)$, which cancels with its square appearing in the denominator and hence leading to a $\phi$ dependence of $\mathcal{T}$ solely via the $\phi$ dependence of energy of ABS ($\omega_0^{\pm}$) appearing in the denominator. Also the $\tau$ dependence of $\mathcal{T}$ enters solely via $\omega_0(\tau,\phi)$ except for the expected overall multiplicative dependence on $\tau$ as shown in R.H.S of Eq.[\ref{4}].  In contrast, for the non-topological JJ, $\mathfrak{f}(\omega,\tau,\phi)=(\omega^2 - \Delta_0^2 \cos \phi -\tau \Delta_0^2 \sin^2 \phi/2)$ \citep{zhao_PRL_91_077003, zhao_PRB_69_134503} which leads to a complicated ($\phi$, $\tau$) dependence of $\mathcal{T}$. We will show the above discussed observations pave the way to identification of distinctive correlation between electrical and thermal currents of a three terminal T-JJ which carry signatures of isolated MBS localized at the junction. \\
\underline{\textit{Signatures of topological ABS in two-terminal thermal} } \underline{\textit{conductance} }: We first discuss signature of the T-JJ pertaining to its $\tau$ dependence in thermal conductance in the simplest case of $\phi=0$. For a non-topological JJ, $\mathcal{T}^{1,2}(\omega,\tau,\phi=0)= 2 \, \tau$ (see Appendix A), thus the thermal conductance of such a JJ in the absence of phase bias is given by (using Eq.\ref{def_of_conductance}) $\kappa= 2\, \tau [G_Q- \int_{0}^{\Delta_0} d\omega\, \omega df(\omega,T)/dT ]_{T=T_{\text{avg}}}$
where $G_Q=\pi^2 k_B^2 T/(3h)$ is quantized thermal conductance of a single ballistic channel and the factor of $2$ represents the doubling due to contributions from particle and hole channels. Hence it equals two times the normal state thermal conductance suppressed up to the gap\citep{zhao_PRL_91_077003, zhao_PRB_69_134503} (see Fig.[\ref{conductance_two_terminal}] (b)). But for the T-JJ, $\mathcal{T}^{1,2}(\omega,\tau,\phi=0)= 2\, \tau (1- \alpha)$ where $\alpha=\Delta_0^2 ((1-\tau) /(\omega^2-\tau \Delta_0^2))$, which implies an additional suppression proportional to $\alpha$ w.r.t. the non-topological JJ. This contrast between topological and non-topological case stem from the fact that at zero phase bias ( $\phi=0$ ) the ABS energy stays pined at $\omega_0^{\pm}=\pm \, \Delta_0$ and hence is independent of transmissivity ($\tau$) of the junction for non-topological case. On the contrary the ABS energy $\vert \omega_0^{\pm} \vert < \Delta_0$ for the topological  case when $\phi=0$ and its exact value depends on $\tau$. Hence this additional suppression of thermal conductance of T-JJ for at $\phi=0$ (see Fig.[\ref{conductance_two_terminal}], (c)) is a direct manifestation of topological ABS. \\ 
\begin{figure}[]
\centering
\includegraphics[scale=0.65]{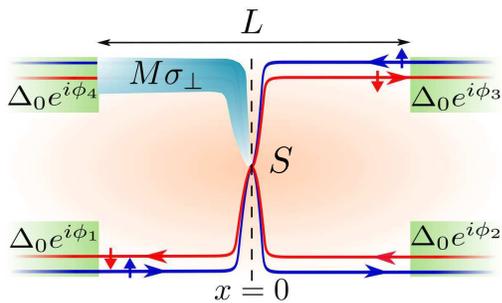}
\caption{Pictorial representation of an effective three terminal JJ depicted on HES of QSH bar geometry where $S$ represent scattering at $x=0$ due to the QPC. $M\sigma_{\perp}$ represents an local magnetic field perpendicular to the spin-quantization axis of HES.}
\label{syatem_diagram}
\end{figure}
Now we start focusing on quantifying the correlation between $\kappa$ and Josephson current discussed above. From (\ref{def_of_conductance}) and (\ref{4}) we note that  the thermal conductance of a T-JJ can be expressed as
\begin{widetext}
\begin{equation}
\kappa = \sigma^N \left[ \dfrac{2}{e^2 \beta(T)} \int_{\Delta_0}^{\infty} d\omega\, \omega \dfrac{ \, (\omega^2-\Delta_0^2)}{(\omega^2 -\vert \omega_0^{\pm} \vert ^2(\tau,\phi))} \dfrac{df(\omega,T)}{dT} \right]_{T=T_{\text{avg}}},
\label{kappa}
\end{equation}
\end{widetext}
where $\sigma^{N}=(e^2\tau/h)\beta(T)$ is Landauer-Buttiker conductance \cite{Landauer,Moskalets} at finite temperature where $\beta(T)=\int_{-\infty}^{\infty} (-\partial f/\partial \omega)d\omega=(4k_BT)^{-1}\int_{-\infty}^{\infty}\left[ \cosh(\frac{\omega}{2k_BT}) \right]^{-2}d\omega$ is the thermal broadening factor. We also note that the  Josephson current  ($I^J$) at temperature $T$ is given by\cite{beenakker_PRL_67_3836} 
\begin{equation}
I^J = - \dfrac{2e}{\hbar} \dfrac{\partial \vert \omega_0^{\pm} \vert }{\partial \phi} \tanh \left( \dfrac{|\omega_0^{\pm}|}{2k_BT} \right).
\end{equation}
Now it is straightforward to note that the Josephson current can be related to the corresponding thermal conductance of the T-JJ via the relation (see Supplemental Material: Section A)
\begin{widetext}
\begin{align}
\partial_{\phi} \kappa &=  \sigma^N \, I^J \, \left[\dfrac{-h |\omega_0^{\pm}|}{\pi e^3 } \bigg{\{} \frac{1}{\beta(T)}  \coth\left( \dfrac{|\omega_0^{\pm}|}{2k_BT} \right)\bigg{\}} \int_{\Delta_0}^{\infty} d\omega \, \omega \dfrac{\omega^2-\Delta_0^2}{(\omega^2-|\omega_0^{\pm}|^2)^2} \dfrac{df(\omega,T)}{dT} \right]_{T=T_{\text{avg}}},
\label{D_K_with_Ij}
\end{align}
\end{widetext}
which directly exploits the fact that the $\phi$ dependence in both $I^J$ and $\kappa$ stem from $\omega_0$ dependence of $\phi$. We  show next that that the three terminal thermal conductance also follows the same relation between $\partial_{\phi_{kl}} \kappa_{i,j}$ and corresponding $\sigma^N_{i,j}$, $I^J_{i,j}$ ($ i,j=1, 2, 3$) owing to the presence of topologically protected MBS at the tri-junction. This fact leads to the central finding of this letter presented in Eq.[\ref{K_with_s}] and  Eq.[\ref{D_K_with_t_and_Ij}].\\
{\underline{\textit{Three Terminal Topological JJ in HES: }}} We consider a Hall bar (QSHB) geometry hosting a QSH state comprising of two helical states with opposite helicity at its opposite edges subjected to a quantum point contact(QPC) which enables us to design an effective four terminal topological JJ (Fig. [\ref{syatem_diagram}]).We also apply a Zeeman filed pointing perpendicular to the spin-polarization axis (i.e. $z$ axis) along left half of the upper edge as shown in the Fig.[\ref{syatem_diagram}] which open up a mass gap in the edge spectrum and reduces the set-up to a three terminal geometry.\\
Hence a three terminal JJ can be simulated by the following BdG Hamiltonian given by
\begin{equation}
\begin{aligned}
\mathcal{H}_{\eta} =& \, H_{edge} + H _{gap}\\
 =& \, \{ ( \eta v_F \hat{p}_x \sigma_z-\mu)\uptau_z 
+& \Delta(x) (\cos \phi_r \uptau_x -\sin \phi_r \uptau_y) \}\\
+& \{(\eta-1)/2\} \,M(x)\sigma_\perp
\label{5}
\end{aligned}
\end{equation}
subjected to a scattering matrix ($S$) imposed at $x=0$ that generates scattering between upper and lower edge (see Fig.[\ref{syatem_diagram}]). We derived  $S$ starting from a local tunnel Hamiltonian representing intra-edge tunneling at $x=0$ in Supplemental Material (Section B) which is used for performing numerical analysis later on. Here $\eta$ defines the helicity of the edge state: $\eta=1$ for lower edge and $\eta=-1$ for upper edge. Pairing potentials in the four superconducting terminals shown in Fig.[\ref{syatem_diagram}] is given by  $\Delta_0e^{i\phi_r}$ where $[r\in \{ 1,2,3,4 \}]$ and $M(x)=M_0 \Theta(x+L/2)\Theta(-x)$ and $M_0\rightarrow \infty$. Other notations has their usual meaning as described before. Assuming that $S$ is such that its elements  respects, $\vert (S)_{i,j} \vert =\vert (S)_{j,i} \vert$ ($i,j=1,2,3$), the solutions for the sub-gap gap bound state ($\omega<\Delta_0$) reduces to $\omega_0=\, 0, \pm \, \Delta_0 \left( \tau_{12} \cos^2 \frac{\phi_{12}}{2} + \tau_{23} \cos^2 \frac{\phi_{23}}{2} + \tau_{13} \sin^2 \frac{\phi_{13}}{2} \right)^{1/2} = \omega_0^0, \omega_0^{\pm}$ (see Supplemental Material: Section C) where $\phi_{ij}=\phi_j-\phi_i$ \footnote{If $\arg{(S)_{i,j}} \neq\arg{(S)_{j,i}}$ then the relative phase of the scattering matrix leads to a additive contribution to $\phi_{i,j}$ leading to anomalous Josepson current. } and $\tau_{ij}=|(S_0)_{i,j}|^2$. Insight into the form of sub-gap bound state solution can be obtained by considering an effective Majorana Hamiltonian where each superconducting terminal ($i=1,2,3$) is expected to contribute one Majorana to the junction except the forth terminal where the Majorana has been pushed away owing to the Zeeman gap.
The effective majorana Hamiltonian then takes the form,\cite{fu_PRB_79_161408, alicea_RPP_75_076501, houzet_PRB_100_014521}
%\begin{figure}[]
%\centering
%\includegraphics[scale=0.65]{Comp.eps}
%\caption{\aabir{Plot of the ratio $\left(\kappa_{1,3}\sigma^{N}_{1,2}\right)/\left(\kappa_{1,2}\sigma^{N}_{1,3}\right)$  for a three terminal non-topological JJ as function of two independent phase differences $\phi_{12}$ and $\phi_{13}$ for different values of $\lambda$ (a) $\lambda=0.1$ (b) $\lambda=0.33$ (c) $\lambda=0.66$ (d) $\lambda=0.9$. For all the plots we have assumed the temperature to be $k_BT=0.5\Delta_0$. The ratio being equal to $1$ is denoted by the solid black lines.}}
%\label{comp}
%\end{figure}
\begin{equation}
H=\dfrac{i}{2}\sum_{1\leq a \leq b \leq 4} \xi_{ab} \gamma_a \gamma_b,
\label{Majorana_Hamiltonian}
\end{equation}
where $\xi_{ab}=\dfrac{\Delta_0}{2}\sqrt{\tau_{ab}} \cos \left( \frac{\phi_{ab}}{2}-\frac{\chi_{ab}}{2} \right)$ and $\gamma_a$ is the Majorana zero mode operator corresponding  to  terminal $a=1,2,3,4$ . We consider $\chi_{ii}=\chi_{13}=\chi_{24}=\pi$ which incorporates the excess Berry phase due to spin flip scattering at the QPC. 
To  mimic our three terminal situation we take $t_{i4}=t_{4i}=0$ for $i \neq 4$ and hence $\xi_{14}=\xi_{24}=\xi_{34}=0$. 
%We consider a fermion basis given by $c_l=(\gamma_1+i \gamma_2)/2$ and $c_u=(\gamma_3+i\gamma_4)/2$. Then the Majorana Hamiltonian in Eq.[\ref{Majorana_Hamiltonian}] expressed in the basis $\mathcal{C}= (c_l,c_u,c_u^{\dagger},-c_l^{\dagger})$ 
This Majorana Hamiltonian readily provides two zero energy eigenvalues and  the other two eigenvalues being identical to $\omega_0^{\pm}$ given above. Note that three of these eigenstate corresponds to  hybridised $\gamma_1$, $\gamma_2$ and $\gamma_3$  resulting in an MBS that stays pinned to zero energy ($\epsilon=0$) and a pair of  Andreev bound states at energy $\epsilon_{\pm}=\omega_0^{\pm}$ while the fourth zero energy state  corresponds to the MBS which stays isolated from the junction. \\
Now returning back to Hamiltonian in Eq.[\ref{5}], for energies $\omega>\Delta_0$, $\mathcal{T}^{i,j}$, the total quasiparticle transmission probability from terminal $j$ to terminal $i$ is given by (see Supplemental Material: Section D)
\begin{equation}
\begin{aligned}
\mathcal{T}^{i,j}(\omega, \tau_{ij}, \phi_{ij})=& \tau_{ij} \left[ \dfrac{2 (\omega^2-\Delta_0^2)~ \mathfrak{f}(\omega, \tau_{ij}, \phi_{ij})}{\prod_{\omega_0} (\omega-\omega_0(\tau_{ij},\phi_{ij}))^2} \right]   \\
&~~~~~~~~~~~~~~~~~~ \times \Theta(\vert \omega\vert^2-\Delta_0^2),
\label{7}
\end{aligned}
\end{equation}
where $[i,j \in \{ 1,2,3 \}, j>i]$ and $\mathfrak{f}=\omega^2 (\omega^2-\vert \omega_0^\pm\vert ^2)$. The resulting expression of  $\mathcal{T}^{n,m}$ has a perfect congruence with two terminal case (see Eq.[\ref{4}]) due to cancellation of $\omega^2$ in the numerator with $(\omega-\omega_0(\tau_{ij},\phi_{ij}))_{\omega_0=\omega_0^0}^2$ in the denominator owing to the MBS at $\omega_0=\omega_0^0=0$. This similarity immediately validates  Eq.[\ref{kappa}] and Eq.[\ref{D_K_with_Ij}] for the three terminal case also where $\kappa \rightarrow \kappa_{i,j}$,  $\phi \rightarrow \phi_{ij}$, $\sigma^N \rightarrow \sigma_{i,j}^N$ and $I^J  \rightarrow I^J_{i,j}$. This fact results in the following relations given by
\begin{equation}
\dfrac{\kappa_{i,j}}{\kappa_{k,l}}=\dfrac{\sigma^N_{i,j}}{\sigma^N_{k,l}},
\label{K_with_s}
\end{equation} 
 \begin{equation}
\dfrac{\partial_{\phi_{kl}} \kappa_{i,j}}{\partial_{\phi_{pq}} \kappa_{m,n}}=\dfrac{\tau_{ij}I^J_{k,l}}{\tau_{mn}I^J_{p,q}}=\dfrac{\sigma^N_{i,j}I^J_{k,l}}{\sigma^N_{m,n}I^J_{p,q}},
\label{D_K_with_t_and_Ij}
\end{equation} \\
which comprise the central finding of this article. Here $i,j,k,l \in \{1,2,3\}$ and  $j\neq i, l\neq k]$. It is remarkable that the ratios of normal state multi-terminal conductance ($\sigma^N_{i,j}$) equates the ratio of corresponding multi-terminal thermal conductance ($ \kappa_{i,j}$) for a T-JJ thought individually they are complicated functions of $T_{\text{avg}}$, $\omega_0$ and $\Delta_0$. $\sigma^N_{i,j}$ being independent of $\phi_{i,j}$, hence the ratios of $ \kappa_{i,j}$ are also independent of $\phi_{i,j}$ though individually they are periodic functions of $\phi_{i,j}$. This results form the special  $\tau_{i,j}$ dependence of $\mathcal{T}^{i,j}$ which is solely via $\tau_{i,j}$ dependence of $\omega_0(\tau_{i,j}, \phi_{i,j})$ except for an overall multiplicative dependence of $\tau_{i,j}$ (see Eq.[\ref{7}]). This property of $\mathcal{T}^{i,j}$ breaks down form non-topological JJ leading to large violation of Eq.[\ref{K_with_s}] which is studied in Supplemental Material: Section F. Even for T-JJ,  violation of Eq.[\ref{K_with_s}] can be observed as we deviate form the short junction limit. A quantitative numerical study on the short to long junction cross-over and the associated violation of Eq.[\ref{K_with_s}] is also presented in Supplemental Material: Section F. Now regarding Eq.[\ref{D_K_with_t_and_Ij}], its is a relation which connects three independent measurable physical quantities, the thermal conductance, the  Josephson current  and the normal state electrical conductance. Similar to Eq.[\ref{K_with_s}], violation of Eq.[\ref{D_K_with_t_and_Ij}] in case of non-topological JJ in the shot junction limit and for T-JJ in case of long junction limit is also studied numerically and is discussed in Supplemental Material: Section F. It should be noted that this kind of relation is unprecedented in studies of JJ and it provides an unique opportunity for putting together outcomes of thermal and electrical measurements to confirm its connection to MBS and hence to its topological origin. In fact both Eq.[\ref{K_with_s}]  and Eq.[\ref{D_K_with_t_and_Ij}]  owe their existence to the presence of an MBS at the tri-junction as discussed below Eq.[\ref{7}] and hence experimental confirmation of these relation can be considered as hallmark of MBS. At the end we would like to reemphasis that though our results  (Eq.[\ref{K_with_s}]  and Eq.[\ref{D_K_with_t_and_Ij}) are derived for a T-JJ hosted in a HES but they are also valid for the case of 1-D electrons with quadratic dispersion and proximitized p-wave superconductivity. This is demonstrated in Supplemental Material: Section G.

\underline{\textit{Acknowledgements}}: A.M. acknowledges Ministry of Human Resource and Development, India for funding. S.D. would like to acknowledge 
the ARF grant from IISER Kolkata.
%
%%%%%%%%%%%%%%%%%%%%%%%%%%%%%%%%%%%%%%%%%%%%%%%%%%%%%%%%%%%%%%%%%%%%%%%%%%%%%%%%%%%%%%%%%%%%%%%%%%%%%%%%%%%%%%%%%%%%%%%%%%

%~~~~~~~~~~~~~~~~~~~~~~~~~~~~~~~~~~~~~~~~~~~~~

%\input{./prob_analysis}

\onecolumngrid

%\section{Appendices}
\appendix
\setcounter{equation}{0}
\setcounter{section}{0}
\renewcommand{\thesection}{\Alph{section}}
\renewcommand{\theequation}{\thesection\arabic{equation}}

\section*{Supplemental Material}

Throughout the calculations we shall assume, unless otherwise mentioned, that the length of the junctions ($L$) are small compared with the superconducting coherence length ($\xi=(\hbar v_F)/\Delta_0$) i.e. $L<<\xi$; $v_F$ being the Fermi velocity. Also we shall focus on the highly doped regime where the chemical potential $\mu>>\Delta_0,k_BT_{\text{avg}}$; $T_{\text{avg}}$ being the average temperature of the junctions.

%%%%%%%%%%%%%%%%%%%%%%%%%%%%% 

\section{Two Terminal Josephson Junctions}
\label{Appendix_A}
\subsection{Bound state energy and total quasiparticle transmission probability in topological Josephson junctions}

We start with the Bogoliubov-de Gennes (BdG Hamiltonian of a two-terminal Josephson junction based on helical edge states, as given in Eq. (3) in the main text
\begin{align}
\mathcal{H} = (-i\hbar v_F \partial_x \sigma_z-\mu)\uptau_z 
+ \Delta(x) (\cos \phi_r \uptau_x -\sin \phi_r \uptau_y),
\label{general_Hamiltonian}
\end{align}
($\Delta(x)=\Delta_0 [\Theta(-x-L/2)+\Theta(x-L/2)]$), $\sigma_i$ and $\uptau_i$ are the Pauli matrices representing spin and particle-hole degrees of freedom respectively and we assume a scattering matrix (for electron) at $x=0$ is given by 
\begin{equation}
S^e =
\begin{pmatrix}
r_{11}	&t	\\
t	&r_{22}
\end{pmatrix}
\label{S_e_2_term}
\end{equation}
The scattering matrix $S^e$ is taken to be symmetric which ensures that there is no time reversal breaking phase which can lead to anomalous Josephson effect.  For energy $\omega<\Delta_0$, bound state energies can be obtained from the equation 
\begin{equation}
\det[\mathbb{I}-a^2(\omega)S^e e^{i \phi}S^h e^{-i\phi}]=0
\end{equation}
where the scattering matrix for hole is given by 
\begin{equation}
S^h=
\begin{pmatrix}
-r_{11}^*	&t^*	\\
t^*	&-r_{22}^*
\end{pmatrix}
\label{S_h_2_term}
\end{equation}
and $a(\omega)=\left( \frac{\omega}{\Delta_0}-i\frac{\sqrt{\Delta_0^2-\omega^2}}{\Delta_0} \right)$ and $e^{i\phi}$ is the diagonal matrix with diagonal elements $\{ e^{i\phi_1},e^{i\phi_2} \}$. Hence the bound state energies are given by  \cite{fu_PRB_79_161408}
\begin{equation}
\omega_0^{\pm}=\pm \Delta_0 \sqrt{\tau} \cos \phi/2
\label{two_term_MBS}
\end{equation}
where $\phi=\phi_2-\phi_1$ and $\tau=|t|^2$.

For energies $\omega>\Delta_0$, we calculate the total quasi-particle transmission probability $(\mathcal{T}^{i,j})$ from terminal $i$ to terminal $j$ $(i,j \in \{ 1,2 \}\,\&\,i\neq j)$. We first consider an electron-like quasiparticle incident on the superconducting lead 1. It will give rise to a reflected electron-like and hole-like quasiparticle within the same lead with amplitudes, say, $r_{ee}$ and $r_{he}$ respectively and transmitted electron-like and hole-like quasiparticle in superconducting lead 2 with amplitudes, say, $t_{ee}^{2,1}$ and $t_{he}^{2,1}$ respectively. The BdG wavefunctions can be written as
%\begin{widetext}
\begin{align}
\Psi_{S1} &= \exp \left[ i \left(\dfrac{\mu+\sqrt{\omega^2-\Delta_0^2}}{\hbar v_F}\right)x \right]
\begin{pmatrix}
e^{\theta/2}e^{i\phi_1/2}	\\
0	\\
e^{-\theta/2}e^{-i\phi_1/2}	\\
0
\end{pmatrix}
+
r_{ee} \exp \left[- i \left(\dfrac{\mu+\sqrt{\omega^2-\Delta_0^2}}{\hbar v_F}\right)x \right]
\begin{pmatrix}
0	\\
e^{\theta/2}e^{i\phi_1/2}	\\
0	\\
e^{-\theta/2}e^{-i\phi_1/2}	\\
\end{pmatrix}	\nonumber \\
&~~~~~~~~~~~~~~~~~~~~~~~~~~~~~~~~~~~~~~~~~~~~~~~~~~~~~~~~~~~~~~~~~
+
r_{hh} \exp \left[ i \left(\dfrac{\mu-\sqrt{\omega^2-\Delta_0^2}}{\hbar v_F}\right)x \right]
\begin{pmatrix}
e^{-\theta/2}e^{i\phi_1/2}	\\
0	\\
e^{\theta/2}e^{-i\phi_1/2}	\\
0
\end{pmatrix}	\\
\Psi_{S2} &= t_{ee}^{2,1} \exp \left[ i \left(\dfrac{\mu+\sqrt{\omega^2-\Delta_0^2}}{\hbar v_F}\right)x \right]
\begin{pmatrix}
e^{\theta/2}e^{i\phi_2/2}	\\
0	\\
e^{-\theta/2}e^{-i\phi_2/2}	\\
0
\end{pmatrix}
+
t_{he}^{2,1} \exp \left[- i \left(\dfrac{\mu-\sqrt{\omega^2-\Delta_0^2}}{\hbar v_F}\right)x \right]
\begin{pmatrix}
0	\\
e^{-\theta/2}e^{i\phi_2/2}	\\
0	\\
e^{\theta/2}e^{-i\phi_2/2}	
\end{pmatrix}	\\
\Psi_{N(1,2)} &= p_{(1,2)} \exp \left[ i\left(\dfrac{\mu+\omega}{\hbar v_F}\right)x \right]
\begin{pmatrix}
1	\\
0	\\
0	\\
0
\end{pmatrix}
+
q_{(1,2)} \exp \left[ -i\left(\dfrac{\mu+\omega}{\hbar v_F}\right)x \right]
\begin{pmatrix}
0	\\
1	\\
0	\\
0
\end{pmatrix}
+
r_{(1,2)} \exp \left[ i\left(\dfrac{\mu-\omega}{\hbar v_F}\right)x \right]
\begin{pmatrix}
0	\\
0	\\
1	\\
0
\end{pmatrix}	\nonumber	\\
&~~~~~~~~~~~~~~~~~~~~~~~~~~~~~~~~~~~~~~~~~~~~~~~~~~~~~~~~~~~~~~~~~~~~~~~~~~~~~~~~~~~~~~~
+
s_{(1,2)} \exp \left[ -i\left(\dfrac{\mu-\omega}{\hbar v_F}\right)x \right]
\begin{pmatrix}
0	\\
0	\\
0	\\
1
\end{pmatrix}
\end{align}
%\end{widetext}
where $\theta=arccosh (\omega/\Delta_0)$. The subscripts $S(N)i$ denote superconducting (normal) region in the $i$th $(i\in \{1,2\})$ terminal. Note that in Fig.\ref{conductance_two_terminal} (a), if we divide the edge into left and right half by drawing an imaginary line across the $x=0$ point then the right and left half will be labeled as terminal-1 and terminal-2 respectively.
By demanding continuity of the wave functions at the boundaries and assuming that the amplitudes of the incoming and the outgoing waves at $x=0$ are related by the scattering matrices (\ref{S_e_2_term}) and (\ref{S_h_2_term}), we get the values of $t_{ee}^{2,1}$ and $t_{he}^{2,1}$. From these we calculate 
\begin{align}
\mathcal{T}_e^{2,1}=|t_{ee}^{2,1}|^2+|t_{he}^{2,1}|^2 &=\dfrac{\tau (\omega^2-\Delta_0^2)\left(\omega^2-\Delta_0^2 \cos^2 (\phi/2)\right)}{(\omega^2-|\omega_0^{\pm}|^2)^2} + \dfrac{\tau (\omega^2-\Delta_0^2)(1-\tau)\Delta_0^2 \cos^2 (\phi/2)}{(\omega^2-|\omega_0^{\pm}|^2)^2} \nonumber \\
&= \tau \dfrac{(\omega^2 -\Delta_0^2)(\omega^2-|\omega_0^{\pm}|^2)}{(\omega^2-|\omega_0^{\pm}|^2)^2}=\tau \dfrac{(\omega^2 -\Delta_0^2)(\omega^2-|\omega_0^{\pm}|^2)}{\prod_{\omega_0}(\omega-\omega_0)^2}
\label{T_2,1}
\end{align}
where $\phi=\phi_2-\phi_1$ and $\omega_0^{\pm}$ is given by (\ref{two_term_MBS}).

Similarly, for a hole-like quasiparticle incident on the first superconducting lead we can calculate $t_{eh}^{2,1}$ and $t_{hh}^{2,1}$ and can define
\begin{equation}
\mathcal{T}_h^{2,1}=|t_{eh}^{2,1}|^2+|t_{hh}^{2,1}|^2
\end{equation}
Also,  $\mathcal{T}_h^{2,1}=\mathcal{T}_e^{2,1}$ and thus
\begin{equation}
\mathcal{T}^{2,1}=\mathcal{T}_e^{2,1}+\mathcal{T}_h^{2,1}=2\tau \dfrac{(\omega^2 -\Delta_0^2)(\omega^2-|\omega_0^{\pm}|^2)}{\prod_{\omega_0}(\omega-\omega_0)^2} = 2 \tau \dfrac{\omega^2-\Delta_0^2}{\omega^2-|\omega_0^{\pm}|^2}
\label{T_2,1_f}
\end{equation}
Also note that $\mathcal{T}^{2,1}=\mathcal{T}^{1,2}$.

The expression of bound state energy (\ref{two_term_MBS}) and $\mathcal{T}^{2,1}$ (\ref{T_2,1}) are same for a normal two terminal Josephson junction with $p$-wave superconductivity \cite{sothmann_PRB_99_214508}.

\subsection{Total quasiparticle transmission probability for non-topological Josephson junction}
For a non-topological  two terminal Josephson junction with $s$-wave superconductivity \cite{zhao_PRL_91_077003, zhao_PRB_69_134503} we have 

\begin{align}
\mathcal{T}_{ee}^{2,1}&=\tau \dfrac{(\omega^2-\Delta_0^2)(\omega^2-\Delta_0^2 \cos^2 \phi/2)}{[\omega^2-\Delta_0^2(1-\tau \sin^2 \phi/2)]^2}	\\
\mathcal{T}_{he}^{2,1}&=\tau(1-\tau) \dfrac{(\omega^2-\Delta_0^2)(\Delta_0^2 \sin^2 \phi/2)}{[\omega^2-\Delta_0^2(1-\tau \sin^2 \phi/2)]^2}
\end{align}
and $\mathcal{T}_e^{2,1}=\mathcal{T}_{ee}^{2,1}+\mathcal{T}_{he}^{2,1}$. Note that, at $\phi=0$, $\mathcal{T}_{he}^{2,1}=0$ and $\mathcal{T}_{ee}^{2,1}=\tau$ so that $\mathcal{T}_e^{2,1}(\phi=0)=\tau$. We also have $\mathcal{T}_h^{2,1}=\mathcal{T}_e^{2,1}$ and thus at $\phi=0$, $\mathcal{T}^{2,1}(=\mathcal{T}_e^{2,1}+\mathcal{T}_h^{2,1})=2\tau$.

\subsection{Thermal conductance, Landauer conductance and Josephson current of a topological Josephson junction}
Single channel spinless normal state electrical conductance with transmission probability $\tau$ is given by Landauer formula \cite{landauer_IBMJRD_1_223, beenakker_PRB_46_12841, Moskalets}
\begin{equation}
\sigma^{N}=\dfrac{e^2 \tau}{h} \beta(T)=\dfrac{e^2 \tau}{h} \int_{-\infty}^{\infty} \left( -\dfrac{\partial f(\omega,T)}{\partial \omega} \right)d\omega=\dfrac{e^2 \tau}{h} (4k_BT)^{-1} \int_{-\infty}^{\infty} \left[ \cosh (\frac{\omega}{2k_BT}) \right]^{-2}d\omega
\label{Landauer_conductance}
\end{equation}
where $e$ is the electronic charge and $f(\omega,T)$ is the Fermi distribution function at temperature $T$. Now, the thermal conductance can be expressed as [Eq.(2) in the main text]
\begin{equation}
\kappa(\phi)=\left[\, \dfrac{1}{h}\int_{\Delta_0}^{\infty} d\omega\, \omega \, \{ \mathcal{T}^{2,1} \} \dfrac{\partial f(\omega,T)}{\partial T}\right]_{T=T_{\text{avg}}}\label{def_of_conductance}
\end{equation}
where $T_{\text{avg}}$ is the average junction temperature.

Now using (\ref{T_2,1_f})
\begin{align}
\kappa &= \dfrac{2\tau}{h} \left[\int_{\Delta_0}^{\infty} d\omega \, \omega \dfrac{\omega^2-\Delta_0^2}{\omega^2-|\omega_0^{\pm}|^2} \dfrac{\partial f(\omega,T)}{\partial T} \right]_{T=T_{\text{avg}}}	\nonumber \\
&= \sigma^N \left[\dfrac{2}{e^2\beta(T)} \int_{\Delta_0}^{\infty} d\omega \, \omega \dfrac{\omega^2-\Delta_0^2}{\omega^2-|\omega_0^{\pm}|^2} \dfrac{\partial f(\omega,T)}{\partial T} \right]_{T=T_{\text{avg}}}
\end{align}
where we have used the expression (\ref{Landauer_conductance}).

Again we know, the Josephson current (at temperature $T$) is given by \cite{beenakker_PRL_67_3836}
\begin{align}
&I^J=I_1+I_2+I_3	\\
&I_1 = - \dfrac{2e}{\hbar} \dfrac{\partial \vert \omega_0^{\pm} \vert }{\partial \phi} \tanh \left( \dfrac{|\omega_0^{\pm}|}{2k_BT} \right)	\\
&I_2 = -\dfrac{2e}{\hbar}2k_BT \int_{\Delta_0}^{\infty} d\omega\, ln[\cosh(\omega/2k_BT)]\dfrac{\partial\rho(\omega,\phi)}{\partial\phi}	\\
&I_3 = \dfrac{2e}{\hbar} \dfrac{d}{d\phi}\int d\vec{r} |\Delta|^2/|g|
\end{align}
where $g$ is the interaction constant of BCS theory. here we have taken $\Delta$ to be independent of $\phi$ so $I_3$ vanishes. $I_2$ is the contribution from the density of states $\rho(\omega,\phi)$ above the gap $\Delta_0$, which also vanishes for short junction limit. Thus, $I^J=I_1$, i.e., the contribution from bound states alone. Now,
\begin{align}
\dfrac{\partial \kappa}{\partial \phi} &= \dfrac{2\sigma^N}{e^2\beta(T)}\dfrac{\partial}{\partial \phi} \left[ \int_{\Delta_0}^{\infty} d\omega \, \omega \dfrac{\omega^2-\Delta_0^2}{\omega^2-|\omega_0^{\pm}|^2} \dfrac{\partial f(\omega,T)}{\partial T} \right]_{T=T_{\text{avg}}}	\nonumber	\\
&= \dfrac{2\sigma^N}{e^2\beta(T)} \left[ \int_{\Delta_0}^{\infty} d\omega \, \omega \dfrac{\omega^2-\Delta_0^2}{(\omega^2-|\omega_0^{\pm}|^2)^2} \dfrac{\partial f(\omega,T)}{\partial T}(-1)(-2 |\omega_0^{\pm}| \dfrac{\partial |\omega_0^{\pm}|}{\partial \phi}) \right]_{T=T_{\text{avg}}}	\nonumber	\\
&= \dfrac{2\sigma^N}{e^2\beta(T)} \left[ \int_{\Delta_0}^{\infty} d\omega \, \omega \dfrac{\omega^2-\Delta_0^2}{(\omega^2-|\omega_0^{\pm}|^2)^2} \dfrac{\partial f(\omega,T)}{\partial T}(2 |\omega_0^{\pm}|)\left(- \dfrac{\hbar I^J}{2e}\coth\left( \dfrac{|\omega_0^{\pm}|}{2k_BT} \right) \right) \right]_{T=T_{\text{avg}}}	\nonumber	\\
&= \sigma^N I^J \left[\dfrac{-h |\omega_0^{\pm}|}{\pi e^3} \left\lbrace \dfrac{1}{\beta(T)}\coth\left( \dfrac{|\omega_0^{\pm}|}{2k_BT} \right)\right\rbrace \int_{\Delta_0}^{\infty} d\omega \, \omega \dfrac{\omega^2-\Delta_0^2}{(\omega^2-|\omega_0^{\pm}|^2)^2} \dfrac{\partial f(\omega,T)}{\partial T} \right]_{T=T_{\text{avg}}}
\end{align}

\section{3 $\times$ 3 scattering matrix from tunnelling Hamiltonian}
\label{Appendix_B}
To derive the scattering matrix at $x=0$, we first write the local Hamiltonian $(\mathcal{H})$ as a sum of edge states Hamiltonian $(\mathcal{H}_{edge})$ and tunnelling Hamiltonian $(\mathcal{H}_T)$ in second quantized notation.
\begin{align}
&\mathcal{H} = \mathcal{H}_{edge}+\mathcal{H}_T	\\
&\mathcal{H}_{edge} = \int dx\{ \psi_{\uparrow}^{\dagger}(x)(-i\hbar v_F \partial_x-\mu)\psi_{\uparrow}(x) +\psi_{\downarrow}^{\dagger}(x)(i\hbar v_F \partial_x-\mu)\psi_{\downarrow}(x)	\nonumber 		\\
&~~~~~~~~~~~~~~~~~~~~~~~~~~~~~~~~~+\psi_{\uparrow}'^{\dagger}(x)(i\hbar v_F \partial_x-\mu)\psi_{\uparrow}'(x)+\psi_{\downarrow}'^{\dagger}(x)(-i\hbar v_F \partial_x-\mu)\psi_{\downarrow}'(x) \}	\\
&\mathcal{H}_T = \hbar v_F \int dx \delta(x) \{ s \, \psi_{\uparrow}^{\dagger}(x)\psi_{\downarrow}(x) + u \, \psi_{\uparrow}'^{\dagger}(x)\psi_{\downarrow}'(x) \nonumber	\\
&~~~~~~~~~~~~~~~~~~~~~~~~~~~~~~~~~~+ t\,(\psi_{\uparrow}^{\dagger}(x)\psi_{\uparrow}'(x)+\psi_{\downarrow}^{\dagger}(x)\psi_{\downarrow}'(x))+v \, (\psi_{\uparrow}^{\dagger}(x)\psi_{\downarrow}'(x)+\psi_{\downarrow}^{\dagger}(x)\psi_{\uparrow}'(x))+h.c.\}
\end{align}
\begin{wrapfigure}{r}{0.5\textwidth}
\begin{center}
\includegraphics[width=0.48\textwidth]{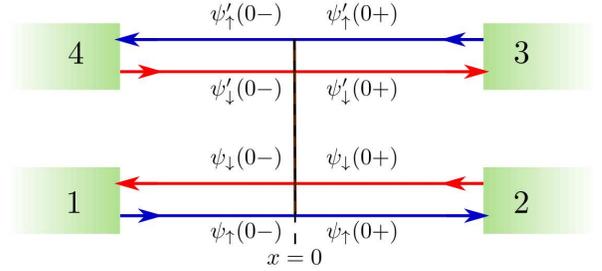}
\end{center}
\caption{Tunnelling between upper and lower edge with different helicity. The blue lines corresponds to up-spin while the red-lines corresponds to down-spin.}
\label{tunnelling_hamiltonian}
\end{wrapfigure}
Here $\psi(x)$ and $\psi'(x)$ represents the lower edge (with helicity $\eta=1$) and the upper edge (with helicity $\eta=-1$). For simplicity, we have assumed the parameters $\{ s,t,u,v \}$ to be real. Apart from the overall factor of $\hbar v_F$; $s$ is the strength of back-reflection in the lower edge while $v$ is that in the upper edge. Inter-edge spin-conserving tunnelling has strength $t$ while that for spin-flipping tunnelling has strength $v$.
We shall follow the standard technique and use the standard Fermionic anticommutation relations \cite{disha_PRB_98_155113}. We note that $\psi_{\uparrow}'(0+)$, $\psi_{\downarrow}'(0-)$, $\psi_{\downarrow}(0+)$ and $\psi_{\uparrow}(0-)$ denote the incoming waves towards the junction, while $\psi_{\uparrow}'(0-)$, $\psi_{\downarrow}'(0+)$, $\psi_{\downarrow}(0-)$ and $\psi_{\uparrow}(0+)$ denote the outgoing waves from the junction [FIG. \ref{tunnelling_hamiltonian}].\\

We represent the scattering matrix as
\begin{equation}
\begin{pmatrix}
\psi_{\downarrow}(0-)	\\
\psi_{\uparrow} (0+)	\\
\psi_{\downarrow}'(0+)	\\
\psi_{\uparrow}' (0-)
\end{pmatrix}
=
\begin{pmatrix}
S_{LR}	&S_{LL}	&S_{LL'}	&S_{LR'}	\\
S_{RR}	&S_{RL}	&S_{RL'}	&S_{RR'}	\\
S_{R'R}	&S_{R'L}	&S_{R'L'}	&S_{R'R'}	\\
S_{L'R}	&S_{L'L}	&S_{L'L'}	&S_{L'R'}
\end{pmatrix}
\begin{pmatrix}
\psi_{\uparrow}(0-)	\\
\psi_{\downarrow} (0+)	\\
\psi_{\uparrow}'(0+)	\\
\psi_{\downarrow}' (0-)
\end{pmatrix}
\end{equation}
which readily defines the terms $S_{ij}$.

Now, using the relation
\begin{equation}
\lim_{\epsilon \to 0} \int_{-\epsilon}^{\epsilon} dx' \{ \psi_{\uparrow}(x),\mathcal{H} \}=0
\end{equation}
we get
\begin{align}
-i(S_{RR}-1)+\dfrac{s}{2}(S_{LR})+\dfrac{t}{2}(S_{L'R})+\dfrac{v}{2} (S_{R'R})&=0	\label{a1}\\
-i(S_{RL})+ \dfrac{s}{2}(1+S_{LL})+\dfrac{t}{2}(S_{L'L})+\dfrac{v}{2}(S_{R'L}) &=0	\label{a2}\\
-i(S_{RL'})+\dfrac{s}{2}(S_{LL'})+\dfrac{t}{2}(1+S_{L'L'})+\dfrac{v}{2}(S_{R'L'}) &=0	\label{a3}\\
-i(S_{RR'}) + \dfrac{s}{2} (S_{LR'}) + \dfrac{t}{2} (S_{L'R'}) +\dfrac{v}{2} (1+S_{R'R'}) &=0	\label{a4}
\end{align}
Similarly, from
\begin{equation}
\lim_{\epsilon \to 0} \int_{-\epsilon}^{\epsilon} dx' \{ \psi_{\downarrow}(x),\mathcal{H} \}=0
\end{equation}
we get,
\begin{align}
i(1-S_{LL})+\dfrac{s}{2}(S_{RL})+\dfrac{t}{2} (S_{R'L})+\dfrac{v}{2} (S_{L'L}) &=0	\label{b1}\\
i(-S_{LR})+\dfrac{s}{2}(S_{RR}+1) +\dfrac{t}{2} (S_{R'R})+\dfrac{v}{2}(S_{L'R}) &=0	\label{b2}\\
i(-S_{LR'})+\dfrac{s}{2}(S_{RR'})+\dfrac{t}{2}(S_{R'R'}+1)+\dfrac{v}{2}(S_{L'R'}) &=0	\label{b3}\\
i(-S_{LL'})+\dfrac{s}{2}(S_{RL'})+\dfrac{t}{2}(S_{R'L'})+\dfrac{v}{2}(S_{L'L'}+1)&=0	\label{b4}
\end{align}
Using
\begin{equation}
\lim_{\epsilon \to 0} \int_{-\epsilon}^{\epsilon} dx' \{ \psi_{\uparrow}'(x),\mathcal{H} \}=0
\end{equation}
we get,
\begin{align}
i(1-S_{L'L'})+\dfrac{u}{2}(S_{R'L'})+\dfrac{t}{2}(S_{RL'})+\dfrac{v}{2}(S_{LL'})&=0	\label{c1}\\
i(-S_{L'R'})+\dfrac{u}{2}(S_{R'R'}+1)+\dfrac{t}{2}(S_{RR'})+\dfrac{v}{2}(S_{LR'}) &=0	\label{c2}\\
i(-S_{L'R})+\dfrac{u}{2}(S_{R'R})+\dfrac{t}{2}(S_{RR}+1)+\dfrac{v}{2}(S_{LR}) &=0	\label{c3}\\
i(-S_{L'L})+\dfrac{u}{2}(S_{R'L})+\dfrac{t}{2}(S_{RL})+\dfrac{v}{2}(S_{LL}+1)	&=0	\label{c4}
\end{align}
Using
\begin{equation}
\lim_{\epsilon \to 0} \int_{-\epsilon}^{\epsilon} dx' \{ \psi_{\downarrow}'(x),\mathcal{H} \}=0
\end{equation}
we get,
\begin{align}
-i(S_{R'R'}-1)+\dfrac{u}{2}(S_{L'R'})+\dfrac{v}{2}(S_{RR'})+\dfrac{t}{2}(S_{LR'}) &=0	\label{d1}\\
-i(S_{R'L'})+\dfrac{u}{2}(1+S_{L'L'})+\dfrac{v}{2}(S_{RL'})+\dfrac{t}{2}(S_{LL'}) &=0	\label{d2}\\
-i(S_{R'R})+\dfrac{u}{2}(S_{L'R})+\dfrac{v}{2}(S_{RR}+1)+\dfrac{t}{2}(S_{LR}) &=0	\label{d3}\\
-i(S_{R'L}) +\dfrac{u}{2}(S_{L'L}) +\dfrac{v}{2} (S_{RL}) +\dfrac{t}{2}(1+S_{LL}) &=0	\label{d4}
\end{align}
Solving the set of equations (\ref{a1})-(\ref{a4}), (\ref{b1})-(\ref{b4}), (\ref{c1})-(\ref{c4}), (\ref{d1})-(\ref{d4}) scattering matrix elements $S_{ij}$ can be determined.

Now, we assume, due to the presence of Zeeman field perpendicular to the spin-polarization axis $(M\sigma_{\perp})$, terminal 4 is totally back-reflecting i.e.
\begin{equation}
\psi_{\uparrow}'(0-)=e^{i\delta}\, \psi_{\downarrow}'(0-)
\end{equation}
where $\delta$ is some phase due to back-reflection. This enables us to write an effective $3\times3$ scattering matrix within the terminals 1, 2 and 3 at $x=0$.
\begin{equation}
\begin{pmatrix}
\psi_{\downarrow}(0-)	\\
\psi_{\uparrow} (0+)	\\
\psi_{\downarrow}'(0+)	\\
\end{pmatrix}
=
\begin{pmatrix}
r_{11}	&t_{12}	&t_{13}	\\
t_{21}	&r_{22}	&t_{23}	\\
t_{31}	&t_{32}	&r_{33}
\end{pmatrix}
\begin{pmatrix}
\psi_{\uparrow}(0-)	\\
\psi_{\downarrow} (0+)	\\
\psi_{\uparrow}'(0+)	\\
\end{pmatrix}
\end{equation}
We define
\begin{equation}
S^e=
\begin{pmatrix}
r_{11}	&t_{12}	&t_{13}	\\
t_{21}	&r_{22}	&t_{23}	\\
t_{31}	&t_{32}	&r_{33}
\end{pmatrix}
\label{ABS_0e}
\end{equation}
Here $r_{ii}$ represents the amplitude of back-reflection within the terminal $i$ and $t_{ji}$ represents the amplitude of transmission from terminal $i$ to terminal $j$. This matrix has the property $t_{ij}=t_{ji}$.

For completeness we write the explicit expressions as calculated,
\begin{small}
\begin{align}
&r_{11}=\dfrac{16i(t^2-su)-4e^{i\delta}[s(4+u^2)-t(tu+4iv)-uv^2]}{e^{i\delta}[4i+s(2-iu)+2u+i(t-v)^2][-4-2iu+s(-2i+u)-(t+v)^2]+4[4u+s^2u-4itv-s(t^2+v^2)]}	\\
&r_{22}=\dfrac{16i(su-v^2)+4e^{i\delta}[s(4+u^2)-t(tu+4iv)-uv^2]}{ie^{i\delta}[-4+2iu+s(2i+u)-(t-v)^2][-4-2iu+s(-2i+u)-(t+v)^2]+4[-4u-s^2u+4itv+s(t^2+v^2)]}
\end{align}
\begin{footnotesize}
\begin{align}
&r_{33}=\dfrac{16+4s^2-8t^2+t^4-2st^2u+4u^2+s^2u^2+8istv-8ituv-8v^2-2t^2v^2-2suv^2+v^4-4ie^{i\delta}[4u+s^2u-4itv-s(t^2+v^2)]}{e^{i\delta}[-4+2iu+s(2i+u)-(t-v)^2][-4-2iu+s(-2i+u)-(t+v)^2]+4i[4u+s^2u-4itv-s(t^2+v^2)]}	
\end{align}
\end{footnotesize}
\begin{align}
&t_{12}=-\dfrac{4i[-4u+s^2u-s(t^2+v^2)]+e^{i\delta}[-4(4+u^2)+s^2(4+u^2)+(t^2-v^2)^2-2s(t^2u+4itv+uv^2)]}{e^{i\delta}[-4+2iu+s(2i+u)-(t-v)^2][-4-2iu+s(-2i+u)-(t+v)^2]+4i[4u+s^2u-4itv-s(t^2+v^2)]}=t_{21}	\\
&t_{13}=\dfrac{-16it+4it^3-4istu-8sv+8uv-4itv^2+4ie^{i\delta}[2it(s+u)+(-4+t^2+su)v-v^3]}{e^{i\delta}[-4+2iu+s(2i+u)-(t-v)^2][-4-2iu+s(-2i+u)-(t+v)^2]+4i[4u+s^2u-4itv-s(t^2+v^2)]}=t_{31}	\\
&t_{23}=\dfrac{8t(-s+u)-4i(4+t^2+su)v+4iv^3-4ie^{i\delta}[t^3-2i(s+u)v-t(-4+su+v^2)]}{e^{i\delta}[-4+2iu+s(2i+u)-(t-v)^2][-4-2iu+s(-2i+u)-(t+v)^2]+4i[4u+s^2u-4itv-s(t^2+v^2)]}=t_{32}
\end{align}
\end{small}
The corresponding scattering matrix for hole $S^h$ can be evaluated by exploiting particle-hole symmetry \cite{calzona_arxiv_1909_06280} and which turns out to be
\begin{equation}
S^h=
\begin{pmatrix}
-r_{11}^*	&t_{12}^*	&-t_{13}^*	\\
t_{21}^*	&-r_{22}^*	&t_{23}^*	\\
-t_{31}^*	&t_{32}^*	&-r_{33}^*
\end{pmatrix}
\label{ABS_0h}
\end{equation}
Particle-hole symmetry Nambu spinor can be understood as follows-
\begin{equation}
\Psi(x)=\sum_{\omega \geq 0} \varphi_{\omega}(x) \gamma_{\omega} +[\mathcal{C}\varphi_{\omega}](x)\gamma_{\omega}^{\dagger}
\end{equation}
where $\Psi(x)=(\psi_{\uparrow(x)},\psi_{\downarrow}(x),\psi_{\downarrow}^{\dagger}(x),-\psi_{\uparrow}^{\dagger}(x))^T$, $\varphi_{\omega}(x)=(u_{\omega,\uparrow}(x),u_{\omega,\downarrow}(x),v_{\omega,\downarrow}(x),v_{\omega,\uparrow}(x))^T$ and the charge conjugation operator $\mathcal{C}=\mathcal{K}\uptau_{z}\otimes \sigma_z$ with $\mathcal{K}$ being the complex conjugation. The operators $\gamma^{\dagger}_{\omega}$ and $\gamma_{\omega}$ are the creation and annihilation operators of a fermionic quasiparticle with energy $\omega$ respectively.

From the fact that BdG Hamiltonian $\mathcal{H}_{\text{BdG}}$ (which is assumed to be diagonal in the basis of $\gamma_{\omega}^{\dagger}$ and $\gamma_{\omega}$) anti-commutes with the charge conjugation operator $\mathcal{C}$ $[\{ \mathcal{H}_{\text{BdG}},\mathcal{C} \}=0]$, it is straight forward to show that if $\varphi_{\omega}$ is the solution of a BdG Hamiltonian $\mathcal{H}_{\text{BdG}}$ with energy $\omega$, then $[\mathcal{C}\varphi_{\omega}]$ will be the solution of the same Hamiltonian $\mathcal{H}_{\text{BdG}}$ with energy $-\omega$. That means upto a global phase we can exploit 
\begin{equation}
\begin{pmatrix}
u_{\omega,\uparrow}(x)	\\
u_{\omega,\downarrow}(x)	\\
v_{\omega,\downarrow} (x)	\\
v_{\omega,\uparrow} (x)
\end{pmatrix}
=
\begin{pmatrix}
-v_{-\omega,\uparrow}^* (x)	\\
v_{-\omega,\downarrow}^* (x)	\\
u_{-\omega,\downarrow}^* (x)	\\
-u_{-\omega,\uparrow}^* (x)
\end{pmatrix}
\label{PH_symmetry}
\end{equation}
This is precisely the particle-hole symmetry.

In this case, we have considered a scattering matrix which is independent of energy $(\omega)$. The full scattering matrix $S$ can be written as
\begin{equation}
S=
\begin{pmatrix}
S^e	&0	\\
0	&S^h
\end{pmatrix}
\end{equation}
and it connects the coefficients $u$ and $v$ as below
\begin{equation}
\begin{pmatrix}
u_{\downarrow}(0-)	\\
u_{\uparrow} (0+)	\\
u_{\downarrow}'(0+)	\\
v_{\downarrow} (0-)	\\
v_{\uparrow} (0+)	\\
v_{\downarrow}' (0+)
\end{pmatrix}
=
\begin{pmatrix}
r_{11}	&t_{12}	&t_{13}	&0	&0	&0	\\
t_{21}	&r_{22}	&t_{23}	&0	&0	&0	\\
t_{31}	&t_{32}	&r_{33}	&0	&0	&0	\\
0	&0	&0	&-r_{11}^*	&t_{12}^*	&-t_{13}^*	\\
0	&0	&0	&t_{21}^*	&-r_{22}^*	&t_{23}^*	\\
0	&0	&0	&-t_{31}^*	&t_{32}^*	&-r_{33}^*
\end{pmatrix}
\begin{pmatrix}
u_{\uparrow} (0-)	\\
u_{\downarrow} (0+)	\\
u_{\uparrow}' (0+)	\\
v_{\uparrow} (0-)	\\
v_{\downarrow} (0+)	\\
v_{\uparrow}' (0+)
\end{pmatrix}
\end{equation}
Note that, we have suppressed the subscript $\omega$ as $S$ is independent of energy. Now under the transformation
\begin{align*}
\begin{pmatrix}
u_{\uparrow}(x)	\\
u_{\downarrow}(x)	\\
v_{\downarrow} (x)	\\
v_{\uparrow} (x)
\end{pmatrix}
\to
\begin{pmatrix}
-v_{\uparrow}^* (x)	\\
v_{\downarrow}^* (x)	\\
u_{\downarrow}^* (x)	\\
-u_{\uparrow}^* (x)
\end{pmatrix}
\end{align*}
[Eq.(\ref{PH_symmetry})] the scattering matrix $S$ remains invariant confirming the particle-hole symmetry of $S$.

\section{Bound state energies of an effective three terminal Josephson junction based on quantum Hall bar geometry\footnote{FIG. 2 of main text}}
\label{Appendix_C}
For energies $\omega<\Delta_0$, the superconducting paring potential, we have decaying solutions in the superconducting leads and propagating solutions in the normal region. Also the terminal 4 plays no role due to the applied Zeeman field. The solutions may be given as
\begin{align}
\Psi_{S1} &=
a_1 \exp \left[ \left( \dfrac{-i\mu+\sqrt{\Delta_0^2-\omega^2}}{\hbar v_F}\right)x \right]
\begin{pmatrix}
0	\\
e^{i\theta/2}e^{i\phi_1/2}	\\
0	\\
e^{-i\theta/2}e^{-i\phi_1/2}	\\
\end{pmatrix} + b_1 \exp \left[ \left( \dfrac{i\mu+\sqrt{\Delta_0^2-\omega^2}}{\hbar v_F} \right)x \right]
\begin{pmatrix}
e^{-i\theta/2}e^{i\phi_1/2}	\\
0	\\
e^{i\theta/2}e^{-i\phi_1/2}	\\
0
\end{pmatrix}	\\
\Psi_{S2} &= a_2 \exp \left[  \left(\dfrac{i\mu-\sqrt{\Delta_0^2-\omega^2}}{\hbar v_F}\right)x \right]
\begin{pmatrix}
e^{i\theta/2}e^{i\phi_2/2}	\\
0	\\
e^{-i\theta/2}e^{-i\phi_2/2}	\\
0
\end{pmatrix}
+
b_2 \exp \left[ \left(\dfrac{-i\mu-\sqrt{\Delta_0^2-\omega^2}}{\hbar v_F}\right)x \right]
\begin{pmatrix}
0	\\
e^{-i\theta/2}e^{i\phi_2/2}	\\
0	\\
e^{i\theta/2}e^{-i\phi_2/2}	
\end{pmatrix}	\\
\Psi_{S3} &= a_3 \exp \left[  \left(\dfrac{i\mu-\sqrt{\Delta_0^2-\omega^2}}{\hbar v_F}\right)x \right]
\begin{pmatrix}
0	\\
e^{i\theta/2}e^{i\phi_3/2}	\\
0	\\
e^{-i\theta/2}e^{-i\phi_3/2}	\\
\end{pmatrix}
+
b_3 \exp \left[ \left(\dfrac{-i\mu-\sqrt{\Delta_0^2-\omega^2}}{\hbar v_F}\right)x \right]
\begin{pmatrix}
e^{-i\theta/2}e^{i\phi_3/2}	\\
0	\\
e^{i\theta/2}e^{-i\phi_3/2}	\\
0
\end{pmatrix}	\\
\Psi_{N(1,2)} &= p_{(1,2)} \exp \left[ i\left(\dfrac{\mu+\omega}{\hbar v_F}\right)x \right]
\begin{pmatrix}
1	\\
0	\\
0	\\
0
\end{pmatrix}
+
q_{(1,2)} \exp \left[ -i\left(\dfrac{\mu+\omega}{\hbar v_F}\right)x \right]
\begin{pmatrix}
0	\\
1	\\
0	\\
0
\end{pmatrix}
+
r_{(1,2)} \exp \left[ i\left(\dfrac{\mu-\omega}{\hbar v_F}\right)x \right]
\begin{pmatrix}
0	\\
0	\\
1	\\
0
\end{pmatrix}	\nonumber	\\
&~~~~~~~~~~~~~~~~~~~~~~~~~~~~~~~~~~~~~~~~~~~~~~~~~~~~~~~~~~~~~~~~~~~~~~~~~~~~~~~~~~~~~~~
+
s_{(1,2)} \exp \left[ -i\left(\dfrac{\mu-\omega}{\hbar v_F}\right)x \right]
\begin{pmatrix}
0	\\
0	\\
0	\\
1
\end{pmatrix}	\\
\Psi_{N3} &= p_{3} \exp \left[ i\left(\dfrac{\mu+\omega}{\hbar v_F}\right)x \right]
\begin{pmatrix}
0	\\
1	\\
0	\\
0
\end{pmatrix}
+
q_{3} \exp \left[ -i\left(\dfrac{\mu+\omega}{\hbar v_F}\right)x \right]
\begin{pmatrix}
1	\\
0	\\
0	\\
0
\end{pmatrix}
+
r_{3} \exp \left[ i\left(\dfrac{\mu-\omega}{\hbar v_F}\right)x \right]
\begin{pmatrix}
0	\\
0	\\
0	\\
1
\end{pmatrix}	\nonumber	\\
&~~~~~~~~~~~~~~~~~~~~~~~~~~~~~~~~~~~~~~~~~~~~~~~~~~~~~~~~~~~~~~~~~~~~~~~~~~~~~~~~~~~~~~~
+
s_{3} \exp \left[ -i\left(\dfrac{\mu-\omega}{\hbar v_F}\right)x \right]
\begin{pmatrix}
0	\\
0	\\
1	\\
0
\end{pmatrix}
\end{align}
where $\theta=\arccos(\omega/\Delta_0)$. The subscripts $S(N)i$ denote superconducting (normal) region in the $i$th $(i\in \{ 1,2,3 \})$ terminal.

By demanding continuity of the wave functions across the boundaries and assuming that the amplitudes of the incoming and outgoing waves at $x=0$ are related by the scattering matrices (\ref{ABS_0e}) and (\ref{ABS_0h}), we get the condition of bound states
\begin{equation}
det[\mathbb{I}-a^2(\omega)S^e e^{i\phi}S^h e^{-i\phi}]=0
\label{BS_three}
\end{equation}
where $a(\omega)=\left( \frac{\omega}{\Delta_0}-i\frac{\sqrt{\Delta_0^2-\omega^2}}{\Delta_0} \right)$ and $e^{i\phi}$ is the diagonal matrix with diagonal elements $\{ e^{i\phi_1},e^{i\phi_2}, e^{i\phi_3} \}$.

Here we note some relations involving the elements of $S$ followed from the unitarity conditions that will be useful in simplifying the algebric expressions. We denote $\tau_{ij}=|t_{ij}|^2$.
\begin{align}
&r_{11}r_{11}^*+\tau_{12}+\tau_{13}=1	\label{B1}	\\
&r_{22}r_{22}^*+\tau_{12}+\tau_{23}=1	\label{B2}	\\
&r_{33}r_{33}^*+\tau_{13}+\tau_{23}=1	\label{B3}	\\
&r_{11}t_{12}^*t_{23}t_{13}^*+r_{11}^*t_{12}t_{23}^*t_{13} = \tau_{12}\tau_{13}-\tau_{13}\tau_{23}-\tau_{12}\tau_{23}	\label{U1}	\\
&r_{22}t_{12}^*t_{13}t_{23}^*+r_{22}^*t_{12}t_{13}^*t_{23} = \tau_{12}\tau_{23}-\tau_{13}\tau_{23}-\tau_{12}\tau_{13}	\label{U2}	\\
&r_{33}t_{13}^*t_{12}t_{23}^*+r_{33}^*t_{13}t_{12}^*t_{23} = \tau_{13}\tau_{23}-\tau_{12}\tau_{13}-\tau_{12}\tau_{23}	\label{U3}	\\
&r_{11}r_{11}^*r_{22}r_{22}^*+r_{11}r_{11}^*r_{33}r_{33}^*+r_{22}r_{22}^*r_{33}r_{33}^*-r_{11}^*r_{22}^*t_{12}^2-r_{11}r_{22}t_{12}^{*2}-r_{11}^*r_{33}^*t_{13}^2-r_{11}r_{33}t_{13}^{*2}-r_{22}^*r_{33}^*t_{23}^2-r_{22}r_{33}t_{23}^{*2}	\nonumber \\
&~~~~~~~~~~~~~~~~~~~~~~~~+t_{12}^2t_{12}^{*2}+t_{13}^2t_{13}^{*2}+t_{23}^2t_{23}^{*2} = 3-2(\tau_{12}+\tau_{13}+\tau_{23})
\label{U4}	\\
&det[S^e] = r_{11}r_{22}r_{33}-r_{33}t_{12}^2-r_{22}t_{13}^2-r_{11}t_{23}^2+2t_{12}t_{13}t_{23}=e^{i\zeta}	\label{U5}	\\
&det[S^h] =-[ r_{11}^*r_{22}^*r_{33}^*-r_{33}^*t_{12}^{*2}-r_{22}^*t_{13}^{*2}-r_{11}^*t_{23}^{*2}+2t_{12}^*t_{13}^*t_{23}^*]=-e^{-i\zeta}	\label{U6}
\end{align}
where $\zeta$ is some arbitrary phase. Using these relations (\ref{B1})-(\ref{U6}) in (\ref{BS_three}) we get the bound state energies $(\omega_0)$ as
\begin{align}
\omega_0^0&=0	\\
\omega_0^{\pm}&=\pm\Delta_0 \sqrt{\tau_{12}\cos^2 \frac{\phi_{12}}{2}+\tau_{13}\sin^2 \frac{\phi_{13}}{2}+\tau_{23}\cos^2\frac{\phi_{23}}{2}}
\end{align}
where $\phi_{ij}=\phi_j-\phi_i$.

\section{Total quasiparticle transmission probability in effective three terminal Josephson junction based on quantum Hall bar\footnote{FIG. 2 of main text}}
\label{Appendix_D}
For energies $\omega>\Delta_0$, we calculate the total quasiparticle transmission probability $(\mathcal{T}^{j,i})$ from terminal $i$ to terminal $j$. We first consider an electron-like quasiparticle incident on the superconducting lead 1. It will give rise to a reflected electron-like and hole-like quasiparticle within the same lead with amplitudes, say, $r_{ee}$ and $r_{he}$ respectively and transmitted electron-like and hole-like quasiparticle in superconducting $n$th lead with amplitudes, say, $t_{ee}^{n,1}$ and $t_{he}^{n,1}$ respectively. The wave functions can be written as
\begin{align}
\Psi_{S1} &= \exp \left[ i \left(\dfrac{\mu+\sqrt{\omega^2-\Delta_0^2}}{\hbar v_F}\right)x \right]
\begin{pmatrix}
e^{\theta/2}e^{i\phi_1/2}	\\
0	\\
e^{-\theta/2}e^{-i\phi_1/2}	\\
0
\end{pmatrix}
+
r_{ee} \exp \left[- i \left(\dfrac{\mu+\sqrt{\omega^2-\Delta_0^2}}{\hbar v_F}\right)x \right]
\begin{pmatrix}
0	\\
e^{\theta/2}e^{i\phi_1/2}	\\
0	\\
e^{-\theta/2}e^{-i\phi_1/2}	\\
\end{pmatrix}	\nonumber \\
&~~~~~~~~~~~~~~~~~~~~~~~~~~~~~~~~~~~~~~~~~~~~~~~~~~~~~~~~~~~~~~~~~
+
r_{hh} \exp \left[ i \left(\dfrac{\mu-\sqrt{\omega^2-\Delta_0^2}}{\hbar v_F}\right)x \right]
\begin{pmatrix}
e^{-\theta/2}e^{i\phi_1/2}	\\
0	\\
e^{\theta/2}e^{-i\phi_1/2}	\\
0
\end{pmatrix}	\\
\Psi_{S2} &= t_{ee}^{2,1} \exp \left[ i \left(\dfrac{\mu+\sqrt{\omega^2-\Delta_0^2}}{\hbar v_F}\right)x \right]
\begin{pmatrix}
e^{\theta/2}e^{i\phi_2/2}	\\
0	\\
e^{-\theta/2}e^{-i\phi_2/2}	\\
0
\end{pmatrix}
+
t_{he}^{2,1} \exp \left[- i \left(\dfrac{\mu-\sqrt{\omega^2-\Delta_0^2}}{\hbar v_F}\right)x \right]
\begin{pmatrix}
0	\\
e^{-\theta/2}e^{i\phi_2/2}	\\
0	\\
e^{\theta/2}e^{-i\phi_2/2}	
\end{pmatrix}	\\
\Psi_{S3} &= t_{ee}^{3,1} \exp \left[  i \left(\dfrac{\mu+\sqrt{\omega^2-\Delta_0^2}}{\hbar v_F}\right)x \right]
\begin{pmatrix}
0	\\
e^{\theta/2}e^{i\phi_3/2}	\\
0	\\
e^{-\theta/2}e^{-i\phi_3/2}	\\
\end{pmatrix}
+
t_{he}^{3,1} \exp \left[ - i \left(\dfrac{\mu-\sqrt{\omega^2-\Delta_0^2}}{\hbar v_F}\right)x \right]
\begin{pmatrix}
e^{-\theta/2}e^{i\phi_3/2}	\\
0	\\
e^{\theta/2}e^{-i\phi_3/2}	\\
0
\end{pmatrix}	\\
\Psi_{N(1,2)} &= p_{(1,2)} \exp \left[ i\left(\dfrac{\mu+\omega}{\hbar v_F}\right)x \right]
\begin{pmatrix}
1	\\
0	\\
0	\\
0
\end{pmatrix}
+
q_{(1,2)} \exp \left[ -i\left(\dfrac{\mu+\omega}{\hbar v_F}\right)x \right]
\begin{pmatrix}
0	\\
1	\\
0	\\
0
\end{pmatrix}
+
r_{(1,2)} \exp \left[ i\left(\dfrac{\mu-\omega}{\hbar v_F}\right)x \right]
\begin{pmatrix}
0	\\
0	\\
1	\\
0
\end{pmatrix}	\nonumber	\\
&~~~~~~~~~~~~~~~~~~~~~~~~~~~~~~~~~~~~~~~~~~~~~~~~~~~~~~~~~~~~~~~~~~~~~~~~~~~~~~~~~~~~~~~
+
s_{(1,2)} \exp \left[ -i\left(\dfrac{\mu-\omega}{\hbar v_F}\right)x \right]
\begin{pmatrix}
0	\\
0	\\
0	\\
1
\end{pmatrix}	\\
\Psi_{N3} &= p_{3} \exp \left[ i\left(\dfrac{\mu+\omega}{\hbar v_F}\right)x \right]
\begin{pmatrix}
0	\\
1	\\
0	\\
0
\end{pmatrix}
+
q_{3} \exp \left[ -i\left(\dfrac{\mu+\omega}{\hbar v_F}\right)x \right]
\begin{pmatrix}
1	\\
0	\\
0	\\
0
\end{pmatrix}
+
r_{3} \exp \left[ i\left(\dfrac{\mu-\omega}{\hbar v_F}\right)x \right]
\begin{pmatrix}
0	\\
0	\\
0	\\
1
\end{pmatrix}	\nonumber	\\
&~~~~~~~~~~~~~~~~~~~~~~~~~~~~~~~~~~~~~~~~~~~~~~~~~~~~~~~~~~~~~~~~~~~~~~~~~~~~~~~~~~~~~~~
+
s_{3} \exp \left[ -i\left(\dfrac{\mu-\omega}{\hbar v_F}\right)x \right]
\begin{pmatrix}
0	\\
0	\\
1	\\
0
\end{pmatrix}
\end{align}
where $\theta=arccosh(\omega/\Delta_0)$. The subscripts $S(N)i$ denote superconducting (normal) region in the $i$th $(i \in \{ 1,2,3 \})$ terminal.

By demanding continuity of the wave functions across the boundaries and assuming that the amplitudes of the incoming and outgoing waves at $x=0$ are connected by the scattering matrices (\ref{ABS_0e}) and (\ref{ABS_0h}), we get the values of $t_{ee}^{2,1}$, $t_{he}^{2,1}$, $t_{ee}^{3,1}$ and $t_{he}^{3,1}$. From these, using the relations (\ref{B1})-(\ref{U6}) we calculate
\begin{align}
&\mathcal{T}_e^{2,1}=|t_{ee}^{2,1}|^2+|t_{he}^{2,1}|^2= \dfrac{\tau_{12}\,\omega^2(\omega^2-\Delta_0^2)(\omega^2-|\omega_0^{\pm}|^2)}{\omega^2(\omega^2-|\omega_0^{\pm}|^2)^2}=\dfrac{\tau_{12}\,\omega^2(\omega^2-\Delta_0^2)(\omega^2-|\omega_0^{\pm}|^2)}{\prod_{\omega_0}(\omega-\omega_0)^2}	\\
&\mathcal{T}_e^{3,1}=|t_{ee}^{3,1}|^2+|t_{he}^{3,1}|^2= \dfrac{\tau_{13}\,\omega^2(\omega^2-\Delta_0^2)(\omega^2-|\omega_0^{\pm}|^2)}{\omega^2(\omega^2-|\omega_0^{\pm}|^2)^2}=\dfrac{\tau_{13}\,\omega^2(\omega^2-\Delta_0^2)(\omega^2-|\omega_0^{\pm}|^2)}{\prod_{\omega_0}(\omega-\omega_0)^2}
\end{align}
where $\omega_0^{\pm}=\pm\Delta_0 \sqrt{\tau_{12}\cos^2 \frac{\phi_{12}}{2}+\tau_{13}\sin^2 \frac{\phi_{13}}{2}+\tau_{23}\cos^2\frac{\phi_{23}}{2}}$ and $(\phi_{ij}=\phi_j-\phi_i)$.

Similarly, for a hole-like quasiparticle incident on the first superconducting lead we can calculate $t_{eh}^{2,1}$, $t_{hh}^{2,1}$, $t_{eh}^{3,1}$ and $t_{hh}^{3,1}$ and can define
\begin{align}
\mathcal{T}_h^{2,1}&=|t_{hh}^{2,1}|^2+|t_{eh}^{2,1}|^2	\\
\mathcal{T}_h^{3,1}&=|t_{hh}^{3,1}|^2+|t_{eh}^{3,1}|^2
\end{align}
It turns out that $\mathcal{T}_h^{2,1}=\mathcal{T}_e^{2,1}$ and $\mathcal{T}_h^{3,1}=\mathcal{T}_e^{3,1}$ and thus
\begin{align}
\mathcal{T}^{2,1}&=\mathcal{T}_h^{2,1}+\mathcal{T}_e^{2,1}=2 \dfrac{\tau_{12}(\omega^2-\Delta_0^2)}{\omega^2-|\omega_0^{\pm}|^2}	\\
\mathcal{T}^{3,1}&=\mathcal{T}_h^{3,1}+\mathcal{T}_e^{3,1}=2 \dfrac{\tau_{13}(\omega^2-\Delta_0^2)}{\omega^2-|\omega_0^{\pm}|^2}
\end{align}
With the same spirit we can calculate $\mathcal{T}^{1,2}$, $\mathcal{T}^{3,2}$, $\mathcal{T}^{1,3}$ and $\mathcal{T}^{2,3}$. Due to assumed reciprocity of the inter-wire resistance $(\tau_{ij}=\tau_{ji})$, $\mathcal{T}^{i,j}=\mathcal{T}^{j,i}$. It can be easily shown that $\mathcal{T}^{i,j}$ can be written in general as
\begin{equation}
\mathcal{T}^{i,j}=2\dfrac{\tau_{ij}\,\omega^2(\omega^2-\Delta_0^2)(\omega^2-|\omega_0^{\pm}|^2)}{\prod_{\omega_0}(\omega-\omega_0)^2}=2 \dfrac{\tau_{ij}(\omega^2-\Delta_0^2)}{\omega^2-|\omega_0^{\pm}|^2}
\end{equation}

\section{Derivation of scattering matrix symmetric between terminal 1 and 2}
\label{F}

\begin{wrapfigure}{r}{0.4\textwidth}
\begin{center}
\includegraphics[width=0.35\textwidth]{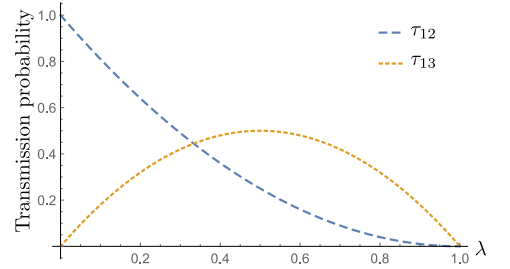}
\end{center}
\caption{Normal state transmission probability from terminal 1 to terminal 2 (and vice versa) $\tau_{12}$ and that from terminal 1 to terminal 3 (and vice versa) $\tau_{13}$ as a function of $\lambda$. $\tau_{12}$ goes faster to zero than $\tau_{13}$ as a function of $\lambda$.}
\end{wrapfigure}

For the scattering matrix discussed in Section \ref{Appendix_B}, we put $s=0$, $u=0$, $\delta=0$, $v=t$ then
\begin{align}
&r_{11}=r_{22}=-1+\dfrac{1}{1+2t^2}=\dfrac{-2t^2}{1+2t^2}=-\left( \dfrac{2t^2}{1+2t^2} \right)	\\
&r_{33}=-1+\dfrac{2}{1+2t^2}=2\left(\dfrac{-2t^2}{1+2t^2}\right)+1=1-2\left( \dfrac{2t^2}{1+2t^2} \right)	\\
&t_{12}=t_{21}=\dfrac{1}{1+2t^2}=\dfrac{1+2t^2-2t^2}{1+2t^2}=1-\left( \dfrac{2t^2}{1+2t^2}\right)	\\
&t_{13}=t_{31}=t_{23}=t_{32}=-\dfrac{2it}{1+2t^2}=-i \left( \dfrac{2t}{1+2t^2}\right)=-i\sqrt{\dfrac{4t^2}{1+2t^2}}	\nonumber	\\
&~~~~~~~~~~~~~~~~~~~~~~~~~~~~~~~~~~~~~~~=-i\sqrt{2\left( \dfrac{2t^2}{1+2t^2}\right)\left(1-\dfrac{2t^2}{1+2t^2}\right)}
\end{align}

Now we note that, for any real number $t$, $0\leq \frac{2t^2}{1+2t^2} \leq 1$. Thus, we redefine $\frac{2t^2}{1+2t^2}=\lambda$, and we have
\begin{align}
&r_{11}=r_{22}=-\lambda	\\
&r_{33}= 1-2\lambda	\\
&t_{12}=t_{21}=1-\lambda	\\
&t_{13}=t_{31}=-i\sqrt{2\lambda(1-\lambda)}=t_{23}=t_{32}
\end{align}
where $0\leq \lambda \leq 1$. Note that, this scattering matrix is symmetric between terminal 1 and terminal 2.

Note that for this scattering matrix $\tau_{12}=(1-\lambda)^2$ and $\tau_{13}=\tau_{23}=2\lambda(1-\lambda)$; so
\begin{equation}
\lim_{\lambda \to 1} \dfrac{\tau_{12}}{\tau_{13}}=\dfrac{\sigma_{1,2}^N}{\sigma_{1,3}^N}=\dfrac{1-\lambda}{2\lambda}=0
\end{equation}
\hspace{10pt}\\

\section{Violation from topological short junction limit}
\label{G}
\hspace{10pt}
\\
\begin{figure}[]
\begin{center}
\includegraphics[scale=0.8]{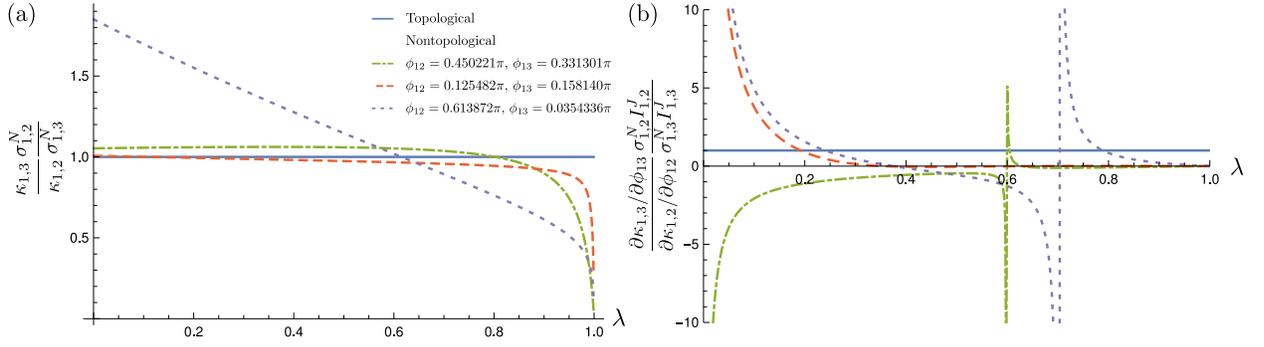}
\end{center}
\caption{The ratios (a) $(\kappa_{1,3}\sigma^N_{1,2})/(\kappa_{1,2}\sigma^N_{1,3})$ and (b) $\left((\partial_{\phi_{13}} \kappa_{1,3}) \sigma^N_{1,2}I^J_{1,2}\right)/\left((\partial_{\phi_{12}} \kappa_{1,2}) \sigma^N_{1,3}I^J_{1,3}\right)$ is $1$ are plotted as a function of scattering matrix parameter $0\leq\lambda\leq1$ for different values of independent phase differences $\phi_{12}$ and $\phi_{13}$. For topological case these ratios are independent of scattering matrix parameters and phase differences unlike non-topological case. The average temperature is assumed to be $k_BT_{\text{avg}}=0.5\Delta_0$}
\label{comp}
\end{figure}

For comparison we assume the same scattering matrix as in Section \ref{F} and check the formulas (11) and (12) as given in the main text. We assume some values of the independent phase differences $\phi_{12}$ and $\phi_{13}$ and plot the ratios $(\kappa_{1,3}\sigma^N_{1,2})/(\kappa_{1,2}\sigma^N_{1,3})$ (FIG. \ref{comp} (a)) and $\left((\partial_{\phi_{13}} \kappa_{1,3}) \sigma^N_{1,2}I^J_{1,2}\right)/\left((\partial_{\phi_{12}} \kappa_{1,2}) \sigma^N_{1,3}I^J_{1,3}\right)$ (FIG. \ref{comp} (b)) as a function of $\lambda$.

For short junction we compare the topological and non-topological three terminal Josephson junctions. For topological case both the ratios are $1$ independent of the values of $\phi_{ij}$ and $(S)_{i,j}$. For non-topological case the ratios are strongly dependent on the phase differences $\phi_{ij}$ and scattering matrix elements $(S)_{i,j}$.

\begin{figure}[]
\begin{center}
\includegraphics[scale=0.8]{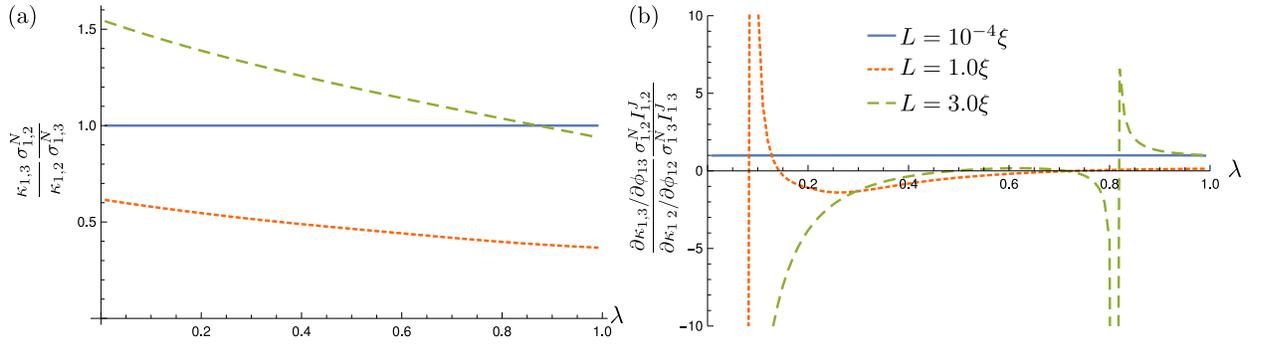}
\end{center}
\caption{The ratios (a) $(\kappa_{1,3}\sigma^N_{1,2})/(\kappa_{1,2}\sigma^N_{1,3})$ and (b) $\left((\partial_{\phi_{13}} \kappa_{1,3}) \sigma^N_{1,2}I^J_{1,2}\right)/\left((\partial_{\phi_{12}} \kappa_{1,2}) \sigma^N_{1,3}I^J_{1,3}\right)$ is $1$ are plotted as a function of scattering matrix parameter $0\leq\lambda\leq1$ for different values of junction length $L$ all for QPC. Here $\xi=\hbar v_f/\Delta_0$ is the superconducting coherence length. Here we have assumed $\phi_{12}=0.880850 \pi$ and $\phi_{13}=0.332947 \pi$. For short junction limit these ratios are independent of scattering matrix parameters and phase differences unlike finite junction limit. The average temperature is assumed to be $k_BT_{\text{avg}}=0.5\Delta_0$.}
\label{long_comp}
\end{figure}

For comparing violation from short junction limit we consider a three terminal Josephson junction based on helical edge states with finite junction length $L$. With the scattering matrix discussed in Section \ref{F} we can calculate the transmission probabilities $\mathcal{T}^{i,j}$ as discussed in Section \ref{Appendix_D}. Now for long junction limit there are two contributions ($I_1$ and $I_2$) to the Josephson current as discussed in Section \ref{Appendix_A}. For accounting both the contribution we shall use the Matsubara sum \cite{brouwer_CSF_8_1249} so the Josephson current is given by
\begin{equation}
I^J_{i,j}=-\dfrac{2e}{\hbar}2k_BT \dfrac{\partial}{\partial \phi_{ij}}\sum_{n=0}^{\infty} \ln \,det [\mathbb{I}-a^2(i\upomega_n)S^e e^{i\phi}S^h e^{-i\phi}]
\end{equation}
where $a(\omega)=\left( \frac{\omega}{\Delta_0}-i\frac{\sqrt{\Delta_0^2-\omega^2}}{\Delta_0} \right)$ and $\upomega_n=(2n+1)\pi k_BT$ are the Matsubara frequencies. We have evaluated the ratios $(\kappa_{1,3}\sigma^N_{1,2})/(\kappa_{1,2}\sigma^N_{1,3})$ (FIG. \ref{long_comp} (a)) and $\left((\partial_{\phi_{13}} \kappa_{1,3}) \sigma^N_{1,2}I^J_{1,2}\right)/\left((\partial_{\phi_{12}} \kappa_{1,2}) \sigma^N_{1,3}I^J_{1,3}\right)$ (FIG. \ref{long_comp} (b)) numerically with $\phi_{12}=0.880850 \pi$ and $\phi_{13}=0.332947 \pi$ for different values of junction length $L$ as a function of scattering matrix parameter $\lambda$. We see that for short junction limit the ratios are $1$ independent of $(S)_{i,j}$ and $\phi_{ij}$.

\begin{figure}[]
\begin{center}
\includegraphics[scale=.65]{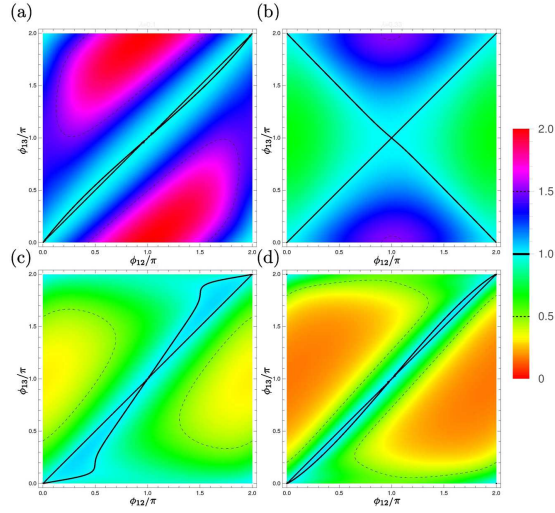}
\end{center}
\caption{Plot of the ratio $\left(\kappa_{1,3}\sigma^{N}_{1,2}\right)/\left(\kappa_{1,2}\sigma^{N}_{1,3}\right)$  for a three terminal non-topological JJ as function of two independent phase differences $\phi_{12}$ and $\phi_{13}$ for different values of $\lambda$ (a) $\lambda=0.1$ (b) $\lambda=0.33$ (c) $\lambda=0.66$ (d) $\lambda=0.9$. For all the plots we have assumed the temperature to be $k_BT=0.5\Delta_0$. The ratio being equal to $1$ is denoted by the solid black lines.}
\label{new_comp}
\end{figure}
A more detailed figure of the ratio $\left(\kappa_{1,3}\sigma^{N}_{1,2}\right)/\left(\kappa_{1,2}\sigma^{N}_{1,3}\right)$ for three terminal non-topological JJ as a function of $\phi_{12}$ and $\phi_{13}$ for different values of $\lambda$ are shown in FIG. \ref{new_comp}.
%\\ \\ \\ \\ \\ 
%\pagebreak

\section{Three terminal normal Josephson junction with $p$-wave superconductivity}
\label{Appendix_E}

\begin{wrapfigure}{l}{0.4\textwidth}
\begin{center}
\includegraphics[width=0.35\textwidth]{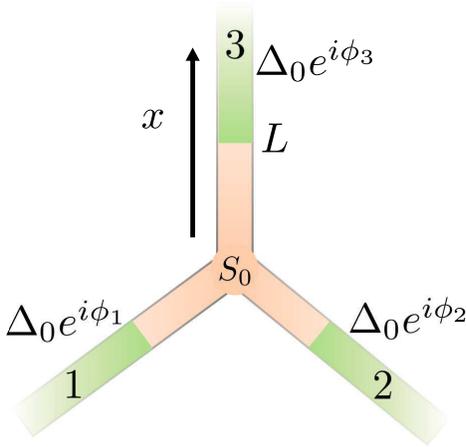}
\end{center}
\caption{Pictorial representation of a three terminal Josephson junction based on normal metal and with $p$-wave superconductivity. The coordinate system is chosen in such a way that $x=0$ at the junction of the three terminals and increases in the direction of superconducting leads.}
\end{wrapfigure}
A Josephson junction based on normal metal with $p$-wave superconductivity also hosts Majorana bound states and therefore are topological in nature. We consider the scattering matrix for electrons $S^e$ at $x=0$ to be of the same form as in (\ref{ABS_0e}). In this case scattering matrix for hole will be $S^h=S^{e*}$. Such junctions can be described using BdG Hamiltonian \cite{thakurathi_JP_27_275702}
\begin{equation}
\mathcal{H}=\left( -\dfrac{\hbar^2}{2m}\dfrac{\partial^2}{\partial x^2}-\mu \right)\uptau_z -i\dfrac{\Delta(x)}{k_F}\dfrac{\partial}{\partial x} [\cos \phi_i \uptau_x-\sin \phi_i \uptau_y]
\end{equation}
where $\uptau_n$ are the Pauli matrices acting on particle-hole basis; $\Delta(x)=\Delta_0 \Theta(x-L)$ and $\phi_i$ is the superconducting phase of the $i$th superconducting lead.

For energies $\omega<\Delta_0$, the solutions in different regions can be written as
\begin{align}
\Psi_{Si}&=a_{i} \exp [i\kappa'_e x]
\begin{pmatrix}
e^{i\theta/2}e^{i\phi_{(2,3)}/2}	\\
e^{-i\theta/2}e^{-i\phi_{(2,3)}/2}
\end{pmatrix}+	\nonumber	\\
&~~~~~~~~~~~~~~~b_{i} \exp [-i\kappa'_h x]
\begin{pmatrix}
e^{-i\theta/2}e^{i\phi_{(2,3)}/2}	\\
-e^{i\theta/2}e^{-i\phi_{(2,3)}/2}
\end{pmatrix}	\\
\Psi_{Ni}&= p_i \exp[ik_e x]
\begin{pmatrix}
1	\\
0
\end{pmatrix}
+
q_i \exp[-ik_e x]
\begin{pmatrix}
1	\\
0
\end{pmatrix}
	\nonumber	\\
	&~~~~~~~~~~~~~~~~+
r_i \exp[ik_h x]
\begin{pmatrix}
0	\\
1
\end{pmatrix}
+
s_i \exp[-ik_h x]
\begin{pmatrix}
0	\\
1
\end{pmatrix}
\end{align}
where $\theta=\arccos(\omega/\Delta_0)$. The subscripts $S(N)i$ denote superconducting (normal) region in the $i$th $(i \in \{ 1,2,3 \})$ terminal. We consider (due to high doping) $\kappa'_e \simeq \kappa'_h \simeq k_e \simeq k_h \simeq k_F=(\sqrt{2m \mu})/\hbar$.

By demanding the continuity of the wave functions and their first derivatives across the boundaries and assuming that the amplitudes of the incoming and outgoing waves are related at $x=0$ by the scattering matrices $S^e$ and $S^h=S^{e*}$ we get the condition for bound state
\begin{equation}
\det[\mathbb{I}-a^2(\omega)S^e(-e^{i\phi})S^he^{-i\phi}]=0
\end{equation}
where $a(\omega)=\left( \frac{\omega}{\Delta_0}-i\frac{\sqrt{\Delta_0^2-\omega^2}}{\Delta_0} \right)$ and $e^{i\phi}$ is the diagonal matrix with diagonal elements $\{ e^{i\phi_1},e^{i\phi_2},e^{i\phi_3} \}$.

This gives the bound state energies $\omega_0$ and are given by $\omega=\omega_0^0,\omega_0^{\pm}$ \cite{oindrila_PRB_97_174518} where
\begin{align}
&\omega_0^0=0	\\
&\omega_0^{\pm}=\pm \Delta_0 \sqrt{\tau_{12} \sin^2 \frac{\phi_{12}}{2}+\tau_{13} \sin^2 \frac{\phi_{13}}{2}+\tau_{23} \sin^2 \frac{\phi_{23}}{2}}
\end{align}
[$\phi_{ij}=\phi_j-\phi_i$; $\tau_{ij}=|t_{ij}|^2$].

For energies $\omega>\Delta_0$, we can calculate the total quasiparticle tunnelling probability $(\mathcal{T}^{i,j})$ from terminal $i$ to terminal $j$. We first consider an electron-like quasiparticle incident on the first superconducting lead. It will give rise to a reflected electron-like and hole-like quasiparticle within the same lead with amplitudes say $r_{ee}$ and $r_{he}$ respectively and transmitted electron-like and hole-like quasiparticle in superconducting $n$th lead with amplitudes, say, $t_{ee}^{n,1}$ and $t_{he}^{n,1}$ respectively. The wave functions can be written as\\

\begin{align}
\Psi_{S1}&= \exp[ -i\kappa_e x]
\begin{pmatrix}
e^{\theta/2} e^{i\phi_1/2}	\\
-e^{-\theta/2} e^{-i\phi_1/2}
\end{pmatrix}
+r_{ee} \exp[ i\kappa_e x ]
\begin{pmatrix}
e^{\theta/2}e^{i\phi_1/2}	\\
e^{-\theta/2}e^{-i\phi_1/2}
\end{pmatrix}
+r_{he} \exp[ -i\kappa_h x ]
\begin{pmatrix}
e^{-\theta/2}e^{i\phi_1/2}	\\
-e^{\theta/2}e^{-i\phi_1/2}
\end{pmatrix}	\\
\Psi_{S(2,3)}&=t_{ee}^{(2,3),1} \exp [i\kappa_e x]
\begin{pmatrix}
e^{\theta/2}e^{i\phi_{(2,3)}/2}	\\
e^{-\theta/2}e^{-i\phi_{(2,3)}/2}
\end{pmatrix}+
t_{he}^{(2,3),1} \exp [-i\kappa_h x]
\begin{pmatrix}
e^{-\theta/2}e^{i\phi_{(2,3)}/2}	\\
-e^{\theta/2}e^{-i\phi_{(2,3)}/2}
\end{pmatrix}	\\
\Psi_{Ni}&= p_i \exp[ik_e x]
\begin{pmatrix}
1	\\
0
\end{pmatrix}
+
q_i \exp[-ik_e x]
\begin{pmatrix}
1	\\
0
\end{pmatrix}
+
r_i \exp[ik_h x]
\begin{pmatrix}
0	\\
1
\end{pmatrix}
+
s_i \exp[-ik_h x]
\begin{pmatrix}
0	\\
1
\end{pmatrix}
\end{align}
where $\theta=arccosh(\omega/\Delta_0)$. The subscripts $S(N)i$ denote superconducting (normal) region in the $i$th $(i \in \{ 1,2,3 \})$ terminal. We consider (due to high doping) $\kappa_e \simeq \kappa_h \simeq k_e \simeq k_h \simeq k_F=(\sqrt{2m \mu})/\hbar$.

By demanding the continuity of the wave functions and their first derivatives across the boundaries and assuming that the amplitudes of the incoming and outgoing waves are related at $x=0$ by the scattering matrices $S^e$ and $S^h=S^{e*}$ we get the values of $t_{ee}^{2,1}$, $t_{he}^{2,1}$, $t_{ee}^{3,1}$ and $t_{he}^{3,1}$. From these, using the relations (\ref{B1})-(\ref{U6}) we calculate
\begin{align}
&\mathcal{T}_e^{2,1}=|t_{ee}^{2,1}|^2+|t_{he}^{2,1}|^2= \dfrac{\tau_{12}\,\omega^2(\omega^2-\Delta_0^2)(\omega^2-|\omega_0^{\pm}|^2)}{\omega^2(\omega^2-|\omega_0^{\pm}|^2)^2}=\dfrac{\tau_{12}\,\omega^2(\omega^2-\Delta_0^2)(\omega^2-|\omega_0^{\pm}|^2)}{\prod_{\omega_0}(\omega-\omega_0)^2}	\\
&\mathcal{T}_e^{3,1}=|t_{ee}^{3,1}|^2+|t_{he}^{3,1}|^2= \dfrac{\tau_{13}\,\omega^2(\omega^2-\Delta_0^2)(\omega^2-|\omega_0^{\pm}|^2)}{\omega^2(\omega^2-|\omega_0^{\pm}|^2)^2}=\dfrac{\tau_{13}\,\omega^2(\omega^2-\Delta_0^2)(\omega^2-|\omega_0^{\pm}|^2)}{\prod_{\omega_0}(\omega-\omega_0)^2}
\end{align}
where $\omega_0^{\pm}=\pm\Delta_0 \sqrt{\tau_{12}\sin^2 \frac{\phi_{12}}{2}+\tau_{13}\sin^2 \frac{\phi_{13}}{2}+\tau_{23}\sin^2\frac{\phi_{23}}{2}}$ and $(\phi_{ij}=\phi_j-\phi_i)$.

Similarly, for a hole-like quasiparticle incident on the first superconducting lead we can calculate $t_{eh}^{2,1}$, $t_{hh}^{2,1}$, $t_{eh}^{3,1}$ and $t_{hh}^{3,1}$ and can define
\begin{align}
\mathcal{T}_h^{2,1}&=|t_{hh}^{2,1}|^2+|t_{eh}^{2,1}|^2	\\
\mathcal{T}_h^{3,1}&=|t_{hh}^{3,1}|^2+|t_{eh}^{3,1}|^2
\end{align}
It turns out that $\mathcal{T}_h^{2,1}=\mathcal{T}_e^{2,1}$ and $\mathcal{T}_h^{3,1}=\mathcal{T}_e^{3,1}$ and thus
\begin{align}
\mathcal{T}^{2,1}&=\mathcal{T}_h^{2,1}+\mathcal{T}_e^{2,1}=2 \dfrac{\tau_{12}(\omega^2-\Delta_0^2)}{\omega^2-|\omega_0^{\pm}|^2}	\\
\mathcal{T}^{3,1}&=\mathcal{T}_h^{3,1}+\mathcal{T}_e^{3,1}=2 \dfrac{\tau_{13}(\omega^2-\Delta_0^2)}{\omega^2-|\omega_0^{\pm}|^2}
\end{align}
With the same spirit we can calculate $\mathcal{T}^{1,2}$, $\mathcal{T}^{3,2}$, $\mathcal{T}^{1,3}$ and $\mathcal{T}^{2,3}$. Due to assumed reciprocity of the inter-wire resistance $(\tau_{ij}=\tau_{ji})$, $\mathcal{T}^{i,j}=\mathcal{T}^{j,i}$. It can be easily shown that $\mathcal{T}^{i,j}$ can be written in general as
\begin{equation}
\mathcal{T}^{i,j}=2\dfrac{\tau_{ij}\,\omega^2(\omega^2-\Delta_0^2)(\omega^2-|\omega_0^{\pm}|^2)}{\prod_{\omega_0}(\omega-\omega_0)^2}=2 \dfrac{\tau_{ij}(\omega^2-\Delta_0^2)}{\omega^2-|\omega_0^{\pm}|^2}
\end{equation}
These expressions are same as that derived in Section \ref{Appendix_D}, which readily justifies the fact that the central results of our paper i.e. Eq. (11) and (12) of main text are not only true for a three terminal Josephson junction based on Dirac type spectrum of helical edge states but also for a three terminal Josephson junction made out of 1-D electrons with quadratic dispersion and proximity induced $p$-wave superconductivity.

\bibliography{Paper1.bib}

%%%%%%%%%%%%%%%%%%%%%%%%%%%%%%%%%%%%%%%%%%%%%%%%%%%%%%%%%%%%%%%%%%%%%%%%%%%%%%%%%%%%%%%%%%%%%%%%%%%%%%%%%%%%%%%%%%%%%%%%%%
%%%%%%%%%%%%%%%%%%%%%%%%%%%%%%%%%%%%%%%%%%%%%%%%%%%%%%%%%%%%%%%%%%%%%%%%%%%%%%%%%%%%%%%%%%%%%%%%%%%%%%%%%%%%%%%%%%%%%%%%%%%%

\end{document}